\documentclass{IEEEtran}
\usepackage[dvips]{epsfig}
\usepackage{amsmath}

\usepackage[oldcommands]{algorithm2e}
\usepackage{graphicx}
\usepackage{subfigure}

\ifCLASSINFOpdf
\else
\fi

\begin{document}
%
\title{Qualitative Action Recognition by Wireless Radio Signals in Human-Machine Systems}

\author{\IEEEauthorblockN{Shaohe Lv, Yong Lu, Mianxiong Dong$^1$, Xiaodong Wang, Yong Dou, and Weihua Zhuang$^2$}

\IEEEauthorblockA{National Laboratory of Parallel and Distributed Processing\\
National University of Defense Technology, Changsha, Hunan, China \\
Email: \{shaohelv, ylu8, xdwang, yongdou\}@nudt.edu.cn \\
$^1$Department of Information and Electronic Engineering\\
Muroran Institute of Technology\\
27-1 Mizumoto-cho, Muroran, Hokkaido, 050-8585, Japan\\
Email: mx.dong@csse.muroran-it.ac.jp\\
$^2$Department of Electrical and Computer Engineering\\
University of Waterloo, Waterloo, Ontario, Canada\\
Email: wzhuang@uwaterloo.ca
}}

\maketitle

\begin{abstract}
Human-machine systems required a deep understanding of human behaviors. Most existing research on action recognition has focused on discriminating between different actions, however, the quality of executing an action has received little attention thus far. In this paper, we study the quality assessment of driving behaviors and present WiQ, a system to assess the quality of actions based on radio signals. This system includes three key components, a deep neural network based learning engine to extract the quality information from the changes of signal strength, a gradient based method to detect the signal boundary for an individual action, and an activity-based fusion policy to improve the recognition performance in a noisy environment. By using the quality information, WiQ can differentiate a triple body status with an accuracy of 97\%, while for identification among 15 drivers, the average accuracy is 88\%. Our results show that, via dedicated analysis of radio signals, a fine-grained action characterization can be achieved, which can facilitate a large variety of applications, such as smart driving assistants.

\end{abstract}

\IEEEpeerreviewmaketitle

\section{INTRODUCTION}
\label{sec:intro}

It is very important to understand fine-grained human behaviors for a human-machine system. The knowledge regarding human behaviors is fundamental for better planning of a Cyber-Physical System (CPS)~\cite{DBLP:conf/infocom/AbdelnasserYH15, DBLP:journals/tmc/GuoCHYZW16, DBLP:journals/thms/YuXYG15}. For example, action monitoring has the potential to support a broad array of applications such as elder or child safety, augmented reality, and person identification. In addition, by observing the behaviors of a person, one can obtain important clues to his intentions. Automatic recognition of activities has emerged as a key research area in human-computer interaction~\cite{DBLP:conf/infocom/AbdelnasserYH15, DBLP:journals/corr/GuoLWYH16}.

While state-of-the-art systems achieve reasonable performance for many action recognition tasks, research thus far mainly focused on recognizing ``which'' action is being performed. It can be more relevant for a specific application to recognize whether this task is being performed correctly or not. There are very limited studies on how to extract additional action characteristics, such as the quality or correctness of the execution of an action~\cite{DBLP:conf/aughuman/VellosoBGUF13}.

In this paper, we study the quality assessment of driving behaviors. A driving system is a typical human-machine system. With the rapid development of automatic driving technology, the driving process requires closer interactions between humans and automobiles (machine) and a careful investigation of the behaviors of the driver~\cite{fatiguereview12}. There are several potential applications for quality assessments of driving behaviors. The first application is driving assistance. According to the quality information, one can classify the driver as a novice or as experienced, and then, for the former, the assistance system can provide advices in complex traffic situation. The second potential application is risk control. It provides an important hint of fatigued driving if a driver repeatedly drives at a low quality level. Additionally, long-term driving quality information is meaningful for the car insurance industry.

We explore a technique for qualitative action recognition based on narrowband radio signals. Currently, fatigue detection systems generally rely on computer vision, on-body sensors or on-vehicle sensors to monitor the behaviors of drivers and detect the driver drowsiness~\cite{DBLP:journals/tvt/JiZL04}. In comparison, a radio-based recognition system is non-intrusive, easy to deploy, and can work well in NLOS (non-line-of-sight) scenarios. Additionally, for old used cars or low-configuration cars, it is much easier to install an radio-based system than a sensor-based system.

It is obvious that quality assessment is much more challenging than action recognition. Qualitative action characterization has thus far only been demonstrated in constrained settings, such as in sports or physical exercises~\cite{DBLP:conf/aughuman/VellosoBGUF13, DBLP:conf/chi/VellosoBG13}. Even with high-resolution cameras and other dedicated sensors, for general activities, a deep understanding of the quality of action execution has not been reached.

There are several technical challenges for quality recognition by radio signals such as modeling the action quality, the method of signal fragments extraction, and how to mitigate the effect of noise and interference. We present WiQ, a radio-based system to assess the action quality by leveraging the changes of radio signal strength. There are three key components in this system:
\begin{itemize}
\item \emph{Deep neural network-based learning}: The quality of action is
characterized by the relative variation (e.g., gradient) of the
received signal strength (RSS). A framework based on the deep neural
network is proposed to learn the quality from the gradient sequence
of the signal strength.
\item \emph{Gradient-based boundary detection}: As the signal strength
can vary sharply at the start and end points of an action, a sudden
gradient change is a strong indicator of the action boundary. A
gradient-based method is proposed to extract the signal fragment for
an individual action.
\item \emph{Activity-based fusion}: Typically, a driving task is
completed by a series of actions, referred to as activity. To mitigate
the effect of surrounding noise, we take an activity as a whole,
fusing the information from all of the actions to derive a sound
observation of the action quality. 
\end{itemize}

We build a proof-of-concept prototype of WiQ with the Universal
Software Radio Peripheral (USRP) platform~\cite{USRP:web} and
evaluate the performance with a driving emulator. To the best of our knowledge, this is the first study in which action quality assessment is performed by using wireless signals and a deep learning method. Our results show that, via dedicated analysis of radio signal features, a fine-grained action characterization can be achieved, which leads to a wide range of potential applications, such as smart driving assistants, smart physical exercise training, and healthcare monitoring.

The rest of this paper is organized as follows. Section~\ref{sec:RelatedWork} provides an overview of related works, and Section~\ref{sec:basicIdea} describes the challenges and basic ideas of our
work. Section~\ref{sec:framework} discusses the design of WiQ.
Section~\ref{sec:performanceevaluation} presents the experimental
results. 
Finally, we conclude this research in Section~\ref{sec:Conclusions}.

\section{RELATED WORK}
\label{sec:RelatedWork}


Action recognition systems generally adopt various techniques such
as computer vision~\cite{DBLP:journals/corr/Wang0T15}, inertial
sensors~\cite{DBLP:conf/chi/CohnMPT12}, ultrasonic~\cite{DBLP:conf/chi/GuptaMPT12}, and infrared electromagnetic
radiation. Recently, we have witnessed the emergence of technologies that can localize a user and track his activities based purely on radio reflections off the person's body~\cite{DBLP:conf/nsdi/AdibKK15, DBLP:conf/nsdi/JoshiBKK15, DBLP:conf/mobicom/WangLCG0L14}. The research has pushed the limits of  radiometric  detection  to  a new level, including motion detection~\cite{DBLP:conf/sigcomm/AdibK13}, gesture recognition~\cite{DBLP:conf/mobicom/PuGGP13}, and localization~\cite{DBLP:conf/nsdi/AdibKKM14}. By exploiting radio signals, one can detect, e.g., motions behind walls and the breathing of a person~\cite{DBLP:conf/huc/MelgarejoZRC14, DBLP:journals/csur/YangZL13, DBLP:conf/infocom/XiZLZTLJ14, DBLP:conf/chi/AdibMKKM15}, or even recognize multiple actions simultaneously~\cite{DBLP:conf/nsdi/AdibKK15}.

In general, there are two major stages in a radio recognition system: feature extraction and classification. Thus far, various signal features have been proposed: energy~\cite{DBLP:journals/tmc/SiggSSJB14}, frequency~\cite{DBLP:conf/mobicom/PuGGP13, DBLP:conf/chi/GuptaMPT12}, temporal
characteristics~\cite{DBLP:conf/nsdi/AdibKK15,
DBLP:conf/nsdi/AdibKKM14}, channel state information~\cite{DBLP:conf/mobicom/WangZZWN14, DBLP:conf/infocom/HanWWN14,
DBLP:conf/huc/MelgarejoZRC14, DBLP:conf/mobicom/WangLCG0L14}, angle characteristics~\cite{DBLP:conf/sigcomm/AdibK13}, etc. The RSS, as an energy feature, is easy to obtain and has been used widely in action recognition~\cite{DBLP:journals/tmc/SiggSSJB14}. In comparison, channel state information (CSI) is a finer-grained feature that can capture human motions effectively~\cite{DBLP:conf/mobicom/WangLCG0L14, DBLP:conf/infocom/HanWWN14}. Doppler shift, as a frequency feature, has been used in gesture recognition~\cite{DBLP:conf/mobicom/PuGGP13}. Time of flight (TOF), as a temporal feature, is used in 3D localization~\cite{DBLP:conf/nsdi/AdibKKM14}. Finally, angle of arrival (AOA) is used in direction inference and object imaging~\cite{DBLP:conf/sensys/HuangNG14}.

There are two major classification methods: 1) fingerprint-based mapping, which takes advantage of machine learning techniques to recognize actions~\cite{DBLP:conf/mobicom/WangZZWN14, DBLP:conf/huc/MelgarejoZRC14,
DBLP:conf/mobicom/WangLCG0L14, DBLP:conf/infocom/HanWWN14}, and 2) geometric mapping, which extracts the distance, direction or other parameters to infer the locations or actions of interest~\cite{DBLP:conf/sigcomm/AdibK13, DBLP:conf/nsdi/AdibKKM14, DBLP:conf/mobicom/PuGGP13}.

While several works have explored how to recognize actions, only a few have addressed the problem of analyzing the action quality. In~\cite{DBLP:conf/chi/VellosoBG13}, a Programming by Demonstration (PbD) method is used to study the action quality in weight lifting exercises through Kinect and Razor inertial measurement units. The sensors in a smart phone are utilized to monitor the quality of exercises on a balance board~\cite{DBLP:conf/percom/MoellerRDKHOP}. Similarly, Wii Fit uses a special balance board to analyze yoga, strength and balance exercises. In addition, driving behavior analysis systems generally rely on computer vision to detect eyelid or head movement, on-body sensors to monitor brain waves or heart rate, or pre-installed instruments to detect steering wheel movements. Biobehavioral characteristics are used to infer a driver's habits or vigilance level~\cite{DBLP:journals/tvt/JiZL04, DBLP:conf/visapp/WangCCF14, fatiguereview12}. While promising, these techniques suffer from limitations, such as physical  contact with drivers (e.g., attaching electrodes), high instrumentation overhead, sensitivity to lighting or the requirement of line-of-sight communication (i.e., a driver wearing eye glasses can pose a serious problem for eye characteristic detection). Different from the existing studies, we focus on assessing the action quality based on radio signals.

\section{BASIC IDEA}
\label{sec:basicIdea}

In this section, we first state the quality recognition problem and then the
design challenges. Afterwards, we discuss the basic idea for
characterizing the quality of actions.

\subsection{Problem statement}

It is critical to understand driving habits for many applications such as driver assistance. We consider two representative tasks: (1) driver identification to determine which driver from a set of candidates performs a driving action; and (2) body status recognition to infer the driver's vigilance level. When a person is inattentive or fatigued, driving actions will be performed in a different manner. That is, the quality of driving actions is changed. It is therefore feasible to monitor the body status by measuring the quality of driving actions.

A careful analysis of the action is required to capture the unique feature of driving. As the driving action is generic, it is insufficient to distinguish the driver or his status by identifying actions alone. In fact, different drivers have different driving styles, e.g., an experienced driver can stop a car smoothly, while a novice may be forced to employ sudden braking.

\begin{figure}
\begin{center}
\includegraphics*[scale=0.186]{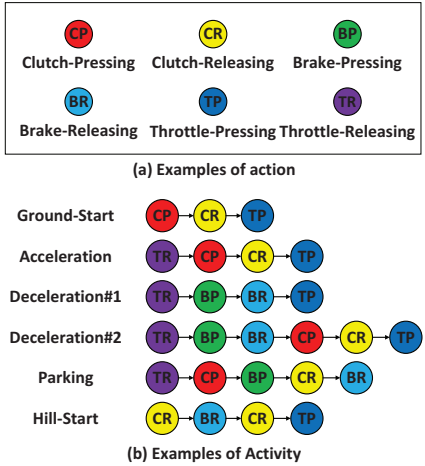}
\end{center}
\caption{ Examples of the actions and typical activities for driving.} \label{fig:fiveActivities}
\end{figure}

In this study, the motions of a driver's foot on the pedal are tracked. Fig.~\ref{fig:fiveActivities}(a) shows six types of actions for manual car driving. Here, an action refers to a short motion that cannot be partitioned further, i.e., a press or release of the pedal (e.g., clutch, brake and throttle). Moreover, an activity is defined as a series of actions to complete a driving task. As shown in Fig.~\ref{fig:fiveActivities}(b), several typical activities are included: ground-start, parking, hill-start, acceleration and deceleration.

To capture the driving behaviors by radio signals, the
transmitter and receiver nodes are located on the two sides of the
pedals. The receiver reports the received signal strength (RSS) per
sampling point. More details about the experimental setup are
described in Section~\ref{sec:performanceevaluation}.


\subsection{Challenges}

There are several technical challenges posed by quality recognition
based on narrowband radio signals.

\textbf{Quality modeling}: Currently, there is no common
understanding regarding what defines the quality of an action. It is
argued that, if one can specify how to perform an action, quality can be defined as the adherence of the execution of an
action to its specification~\cite{DBLP:conf/aughuman/VellosoBGUF13,
DBLP:conf/chi/VellosoBG13}. To measure quality, it is therefore necessary to characterize the execution of an action through a
finer-grained motion analysis. For example, consider the braking
(BP) action in three driving behaviors, e.g., sudden braking,
parking, and slight deceleration. As shown in Table~\ref{table:t1},
though the action is the same, the quality is quite different: (1)
the movement of the brake in the first case is much faster than the
others; and (2) the movement distance of the brake in the last case
is smaller than the others.

\begin{table}[htbp]
 \centering
\renewcommand{\arraystretch}{1.0}
\setlength\tabcolsep{3pt}
  \caption{Comparison of the action quality in three behaviors. }\label{table:t1}

  \begin{tabular}{|llll|}
\hline
\multicolumn{1}{|c}{Quality}  &   \multicolumn{1}{c}{Sudden braking}  &   \multicolumn{1}{c}{Parking}   & \multicolumn{1}{c|}{Slight deceleration}\\
\hline
Speed  &  \multicolumn{1}{c}{fast}  & \multicolumn{1}{c}{regular}  & \multicolumn{1}{c|}{regular} \\
Range  &  \multicolumn{1}{c}{large}  & \multicolumn{1}{c}{large}  & \multicolumn{1}{c|}{small} \\
\hline
\end{tabular}
\end{table}

\begin{figure*}
\begin{minipage}[b]{0.4\linewidth}
\begin{center}
\includegraphics*[scale=0.172]{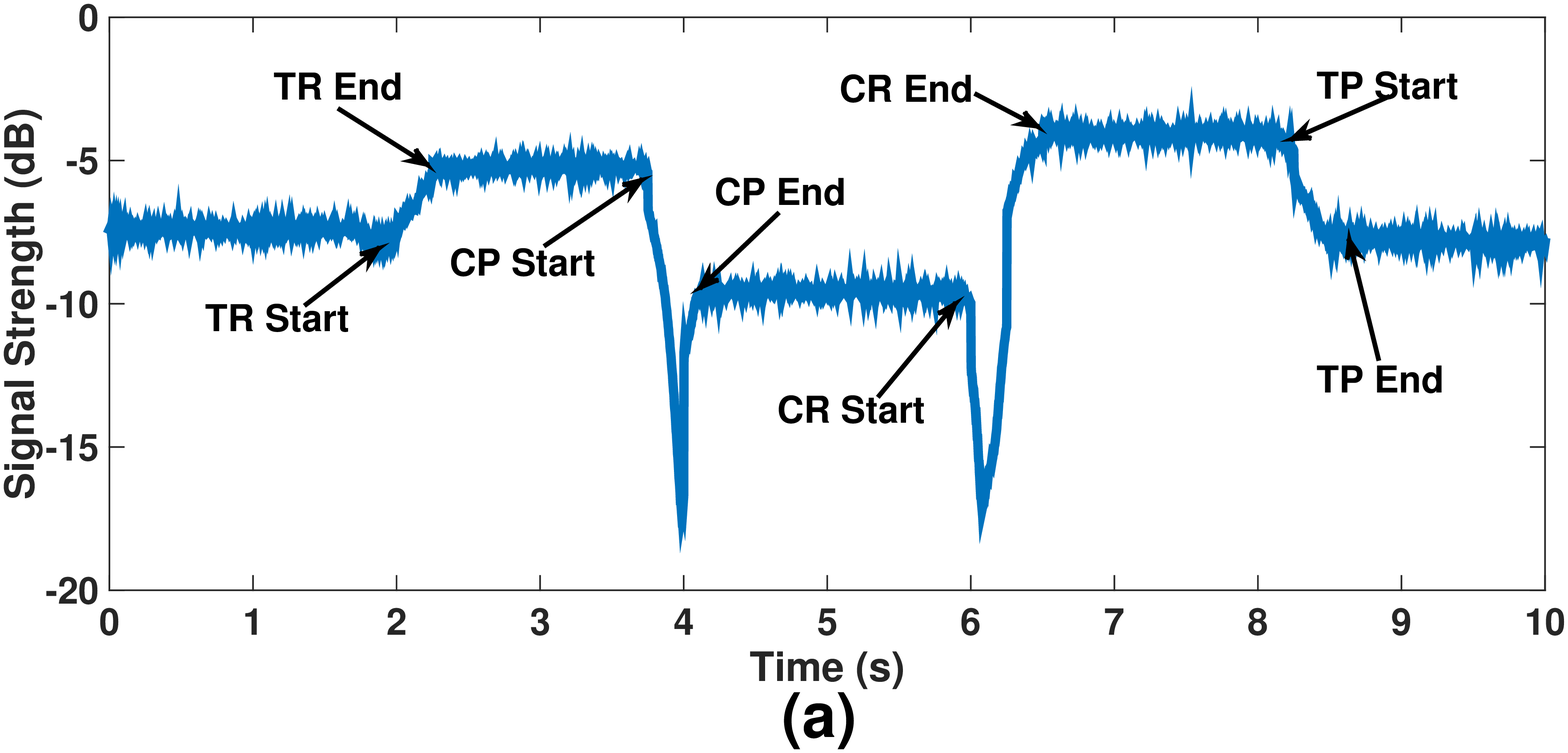}
\end{center}
\end{minipage}
\hspace{0.1\linewidth}
\begin{minipage}[b]{0.4\linewidth}
\begin{center}
\includegraphics*[scale=0.172]{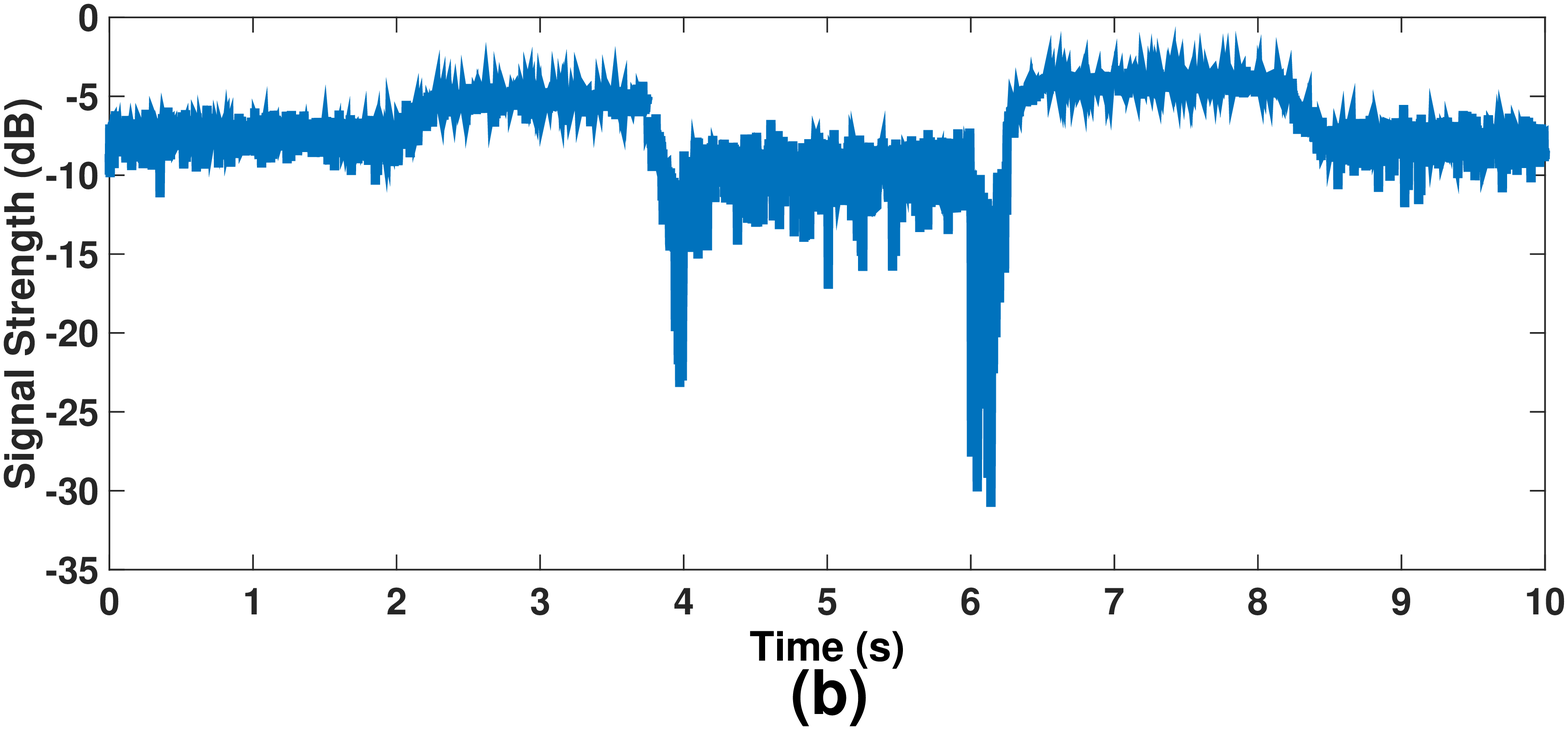}
\end{center}
\end{minipage}
\caption{Received signal strength for the acceleration activity with (a) high
SNR; (b) low SNR.} \label{fig:NoiseEffect}
\end{figure*}

There is currently no effective way to characterize the execution of an action. Though many radio signal features are proposed for action recognition, most of them are used to recognize what types of actions are carried out.

\textbf{Signal fragment extraction}: As the radio signal is
sampled continuously over time, when multiple actions occur sequentially, we
need to partition the signal into several fragments, i.e., one
fragment for one action. As an example, Fig.~\ref{fig:NoiseEffect}(a)
shows the signal for the acceleration activity. To accelerate with a
gear shift, one should release the throttle (TR), press the clutch
(CP) and change the gear (which is invisible here), release the
clutch (CR) and press the throttle (TP) until a desired speed is
reached. To analyze the quality, the start and end points of all
the actions must be identified accurately.

There is no feasible solution to detect the signal boundary. In~\cite{DBLP:conf/mobicom/PuGGP13}, a gradient-based method is used to
partition a Doppler shift sequence. The Doppler shift information is, however, not available in the most modern systems such as in wireless local-area networks (WLANs). A method was recently proposed in
WiGest~\cite{DBLP:conf/infocom/AbdelnasserYH15} to insert a
special preamble to separate different actions, which require interrupting the usual signal processing
routine. Neither can be adopted in our scenarios.

\textbf{Robustness}: Quality assessment can be easily misled by noise or interference in the radio channel. As shown in Fig.~\ref{fig:NoiseEffect}(b), when the signal to noise ratio (SNR) is
low, it is difficult to identify the action and extract the
quality information. Although a denoising method can be used to reduce
the effect of noise or interference, it is necessary to have
an effective way to sense the radio channel condition and
mitigate any negative effect on quality assessment.

\subsection{Quality recognition}

We characterize the quality of action with respect to motion and we consider the \emph{duration} of an execution and the \emph{speed}
and \emph{distance} of the pedal motion. We first discuss the case
of the throttle and then extend our discussion to the clutch and brake.

\begin{figure}
\begin{center}
\includegraphics*[scale=0.152]{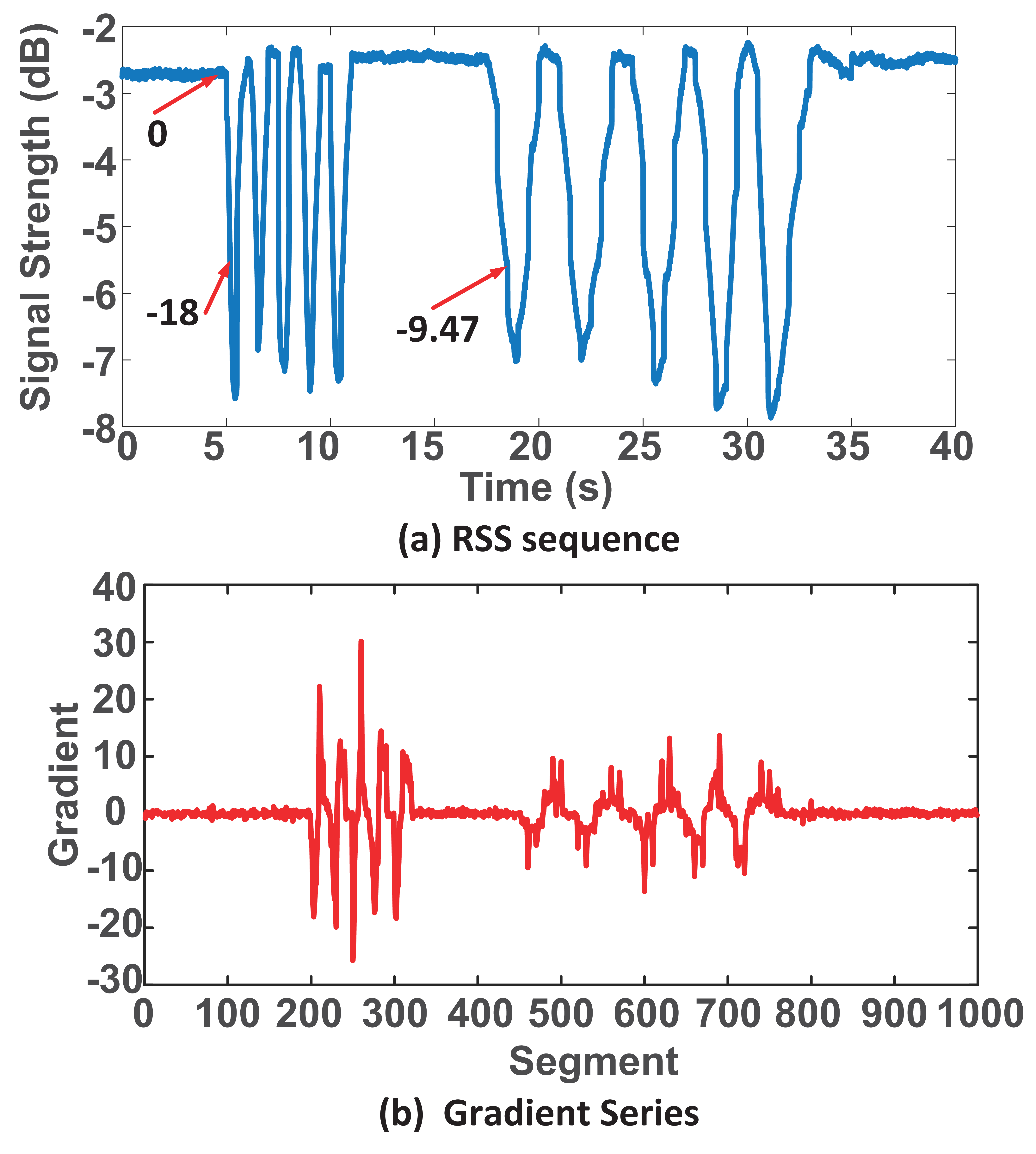}
\end{center}
\caption{The RSS and the gradient when the throttle is pressed and
released, quickly for five times and then slowly for five times.}
\label{fig:acceleratorClap}
\end{figure}

The duration of an execution can be estimated after the signal
boundary of the action is detected. Let $T_S$ be the number of
sampling points in the fragment, an estimate of the duration is
$(T_S-1)\times t_u$, where $t_u$ is the length of the sampling interval.

The movement speed can be captured by the change rate (e.g.,
gradient) of signal strength. Fig.~\ref{fig:acceleratorClap}(a)
shows the received signal in the experiment: the throttle is
pressed and released quickly five times and then slowly another
five times. When the motion is faster, the change of signal strength
is sharper (e.g., the typical gradients are -18 and -9.47 for the two
cases, respectively). The gradient sequence of signal strength is
plotted in Fig.~\ref{fig:acceleratorClap}(b). The gradient
magnitude is, on average, much larger for a quicker motion. Thus, thee
gradient of signal strength is an effective metric to characterize the movement
speed.

\begin{figure}
\begin{center}
\includegraphics*[scale=0.183]{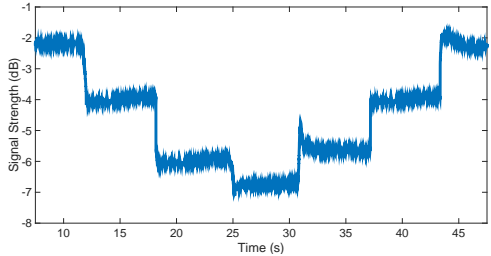}
\end{center}
\caption{Received signal strength for the TP and TR actions when the throttle
is located at different positions.} \label{fig:RSSDistance}
\end{figure}

The correlation between the pedal position and the signal strength
is exploited to estimate the motion distance. We press the throttle
(TP) to a small extent and hold for several seconds; then press it
to a large extent and hold for several seconds and finally press it to
the maximum degree. The same pattern is repeated for the
throttle-releasing (TR) in the opposite order. The received signal strength
is shown in Fig.~\ref{fig:RSSDistance}. The signal strength is
distinct when the pedal position is different. To infer the motion
distance during the action execution, a simple method is to compute the
difference between the signal strengths at the start and end points. 

\begin{figure*}
\begin{minipage}[b]{0.3\linewidth}
\begin{center}
\includegraphics*[scale=0.2]{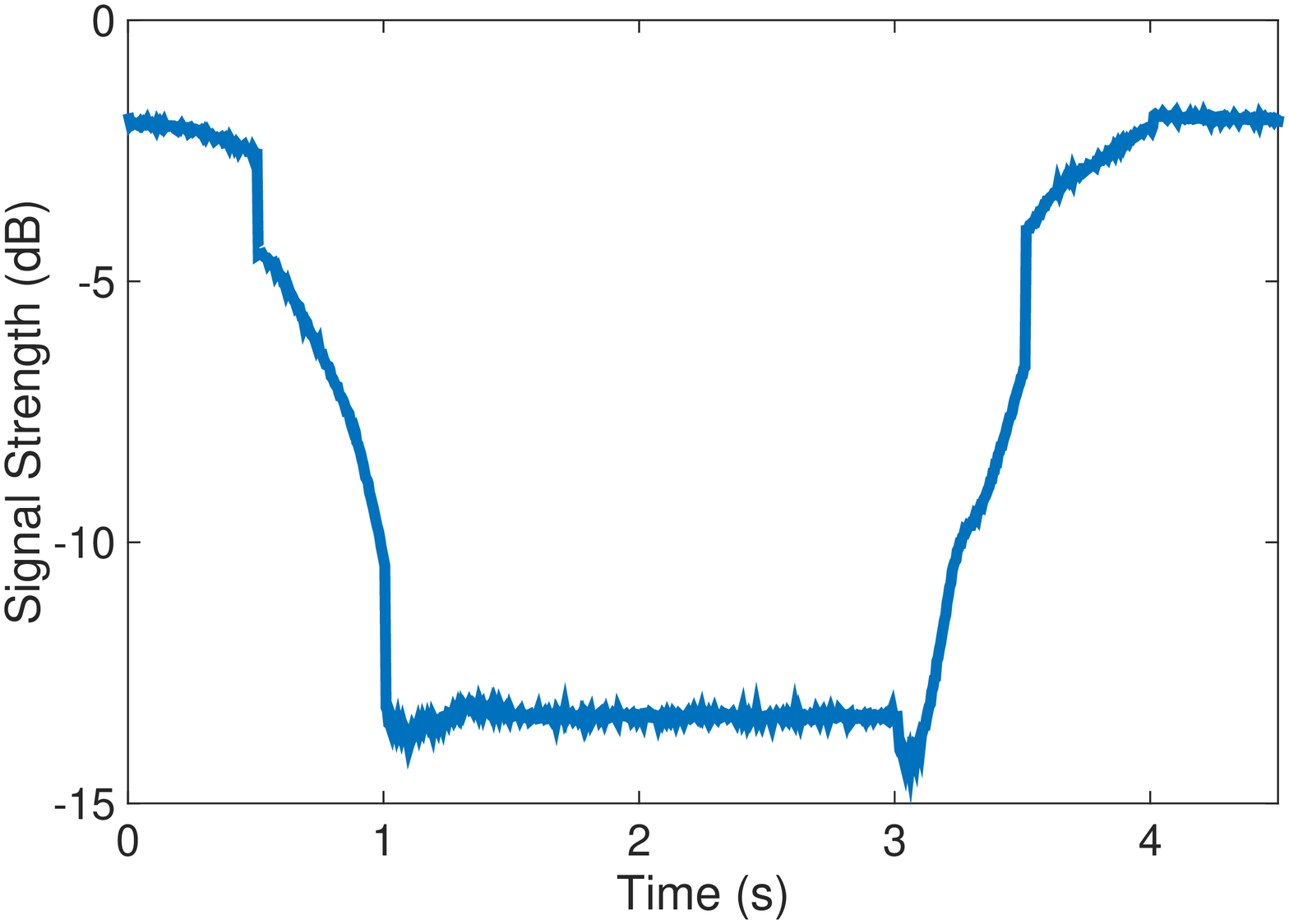}
\end{center}
\end{minipage}
\hspace{0.02\linewidth}
\begin{minipage}[b]{0.32\linewidth}
\begin{center}
\includegraphics*[scale=0.2]{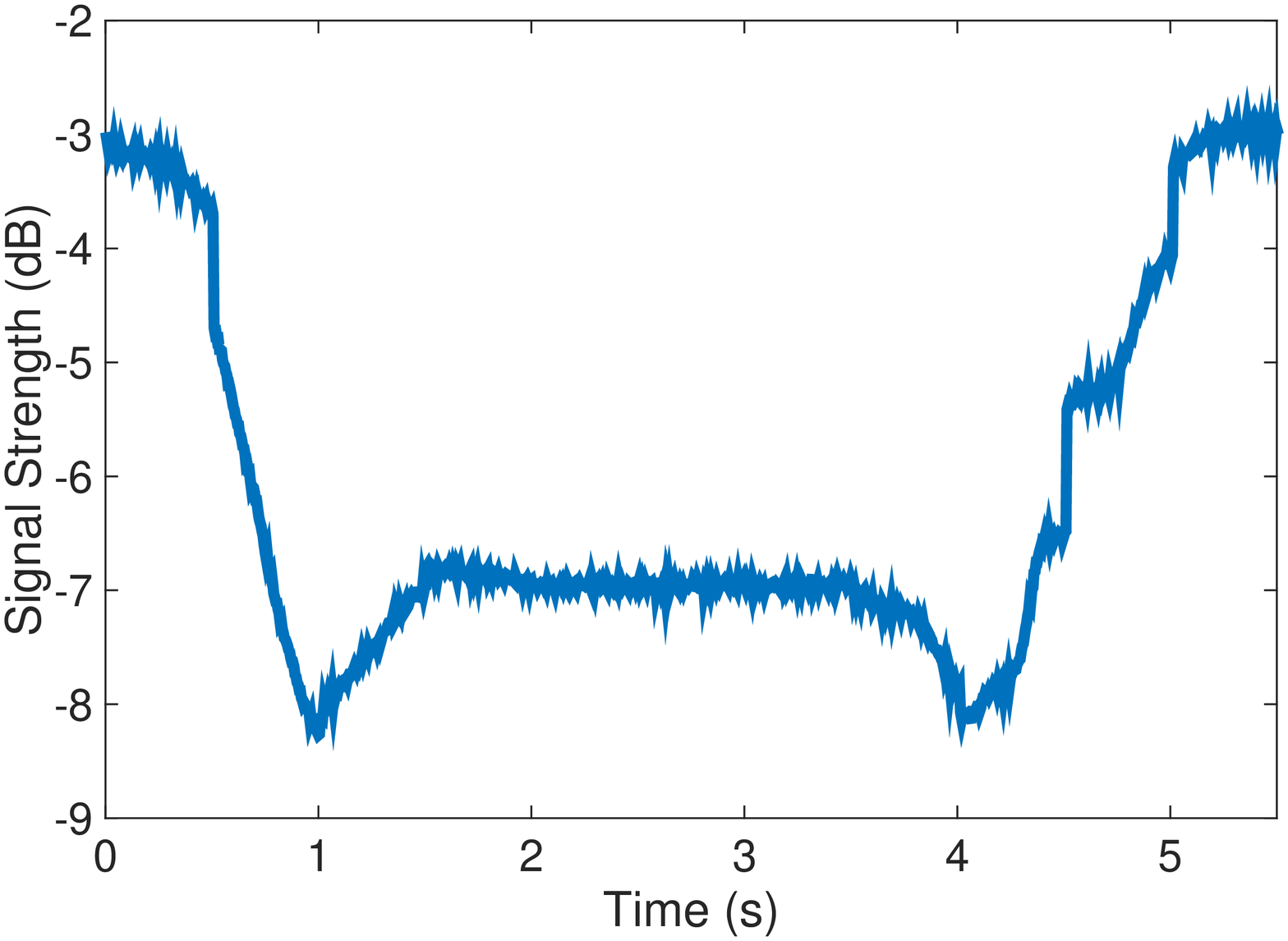}
\end{center}
\end{minipage}
\hspace{0.02\linewidth}
\begin{minipage}[b]{0.32\linewidth}
\begin{center}
\includegraphics*[scale=0.2]{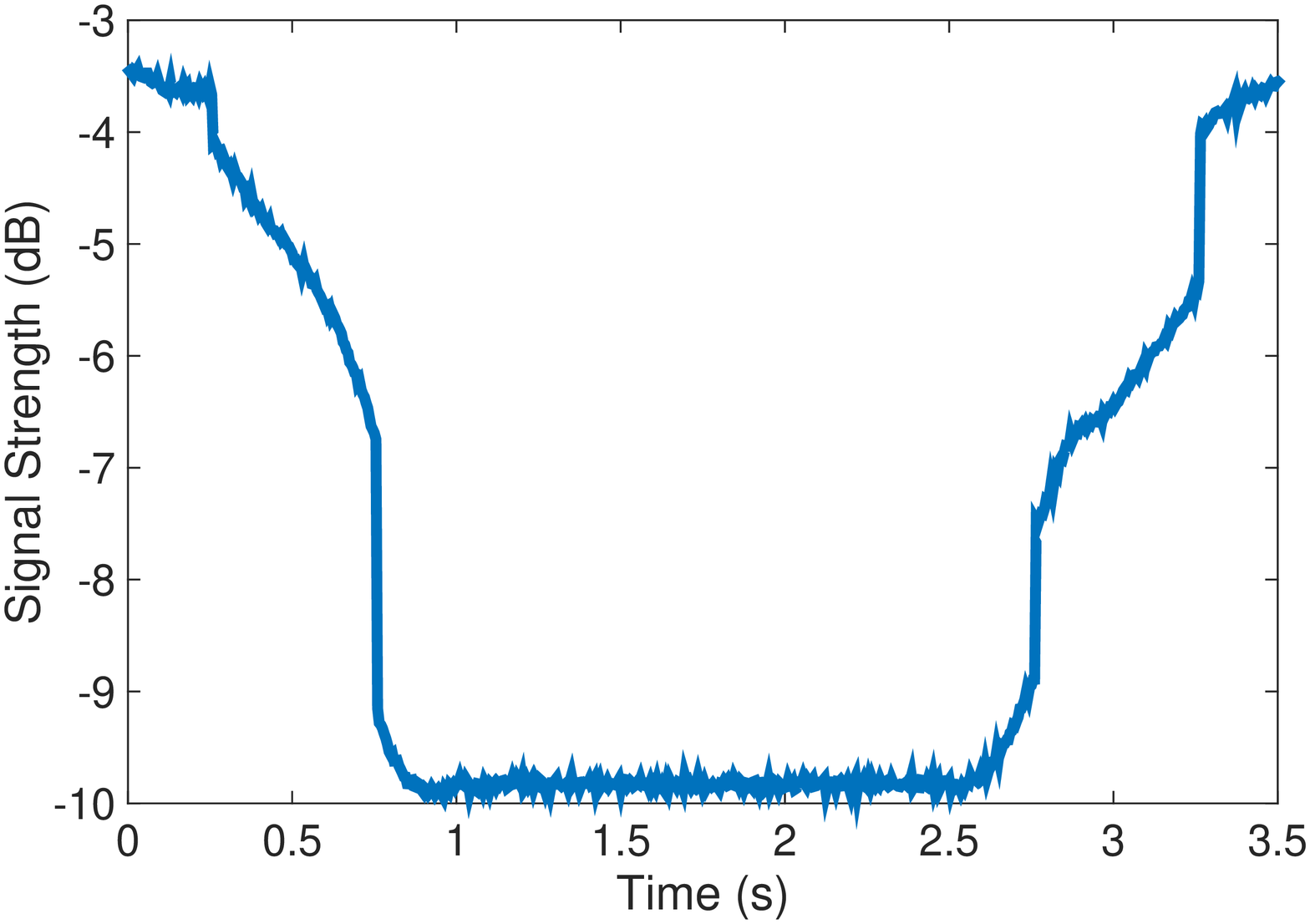}
\end{center}
\end{minipage}
\caption{Received signal strength when the (a) clutch, (b) brake and (c)
throttle are pressed and then released.} \label{fig:RSSAction}
\end{figure*}

For the clutch and brake, it is slightly more complex. As shown in Fig.~\ref{fig:RSSAction}, when the clutch is pressed, the signal strength
first decreases and then increases. The change is no longer
monotonic, which is different from the throttle. A
similar observation can be drawn for the brake. To estimate
the motion distance, we detect the maximal (or minimal) point during
the execution of an action. If one such point is found, letting $S_M$ be
the signal strength, the motion distance can be characterized by the
oscillation range of signal strength, i.e.,
$|S_A-S_M|+|S_M-S_E|$, where $S_A$ and $S_E$ are the
signal strength at two boundary points, respectively. 

As shown in Fig.~\ref{fig:RSSAction}, different patterns of signal
strength can be observed for distinct actions, e.g., the signal
strength decreases consistently during brake-pressing (BP) and
always increases during brake-releasing (BR). One can exploit the
patterns to discriminate among different actions.

\section{DESIGN OF WIQ}
\label{sec:framework}

We first overview the basic procedure of WiQ and then discuss in
detail the three key components, e.g., the learning engine, signal
boundary detection, and decision fusion. We finally discuss some
possible extensions of WiQ.

\subsection{Overview}

WiQ first detects the signal boundary for each action, and then
recognizes the driving action and extracts the motion quality, and
finally identifies the driver or body status.

\begin{figure}
\begin{center}
\includegraphics*[scale=0.302, angle=270]{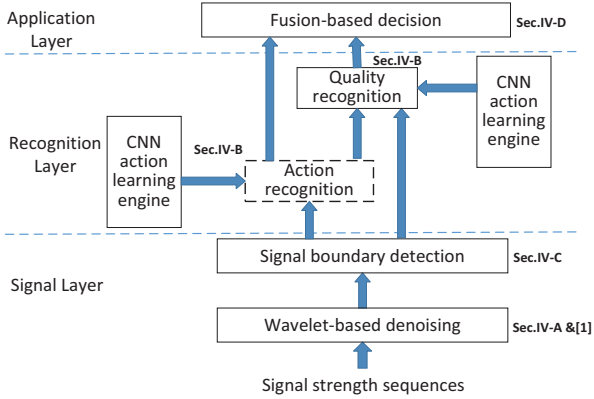}
\end{center}
\caption{Illustration of the basic process of WiQ.}
\label{fig:appOverview}
\end{figure}

Fig.~\ref{fig:appOverview} shows the basic process of WiQ. There are
three layers, i.e., signal, recognition and application. The inputs to the signal layer are the radio signals that capture the driving
behaviors. Due to the complex wireless propagation and interaction
with surrounding objects, the input values are noisy. We leverage a
wavelet-based denoising method to mitigate the effect of the noise
or interference. We here omit the details of the method, which is given in~\cite{DBLP:conf/infocom/AbdelnasserYH15}. Afterwards, a signal
boundary detection algorithm is applied to extract the signal
fragment corresponding to the individual action.

The input of the recognition layer is the fragmented signal for an
action. We first adopt a deep learning method to recognize the
action. Afterwards, the quality of the action is extracted by a deep
learning engine and provided to the upper layer, together with the
results of action recognition.

At the application layer, a classification decision is made.
For driver identification, the \emph{classification} process
determines which driver performs the action. For body status
recognition, the process determines the driver's status according to
the action quality. Additionally, a fusion policy is adopted to improve the
robustness and accuracy.

\subsection{Quality recognition}

There are two major stages in quality recognition: feature
extraction and classification (based on the quality of an action).
In the first stage, we adopt a convolutional neural networks
(CNN). In addition, a normalized multilayer perceptron (NMLP)
is used for classification. Both CNN and NMLP are supervised
machine learning technique~\cite{DBLP:journals/corr/Wang0T15}.

CNN is a representative deep learning method that uses the multilayer neural networks to extract interesting
features. Deep learning, as an effective method of machine learning,
has achieved great success in image recognition, speech
recognition and many other areas~\cite{DBLP:journals/corr/Wang0T15}.
It has been used widely due to its low dependence on
prior-knowledge, small number of parameters and a high training
efficiency.

\begin{figure}
\begin{center}
\includegraphics*[scale=0.3408, angle=270]{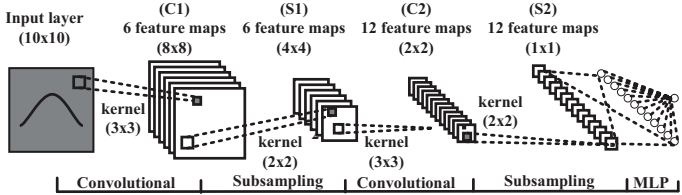}
\end{center}
\caption{A convolutional neural network for quality recognition.} \label{fig:cnnRecProcess}
\end{figure}

To recognize the quality of action, a five-layer CNN network has
been built and the structure is shown in Fig.~\ref{fig:cnnRecProcess}. Basically, there are two convolutional
layers, two sub-sampling layers and one fully-connected layer. In the
first convolutional layer, the size of a convolutional kernel is
$3\times3$. Six different kernels are adopted to generate six
feature maps. At the second convolutional layer, there are two
kernels and the kernel size is still $3\times3$. There are, in total, 12 feature maps as the output of this layer. The goal of a
convolutional layer is to extract as many features as possible in an
effective manner. In comparison, a sub-sampling layer is devoted to combining
the lower-layer features and reducing the data size. There is only one
kernel in the sub-sampling layer and the size is $2\times2$. The last
layer of CNN is a fully-connected layer that combines all the learned
features. The output of the CNN network is a vector of twelve
dimensions, which is the input of the NMLP classifier.

\begin{table}[htbp]
 \centering
\renewcommand{\arraystretch}{2.0}
\setlength\tabcolsep{4pt}
  \caption{Features of gradient for quality recognition. }\label{table:quality}
\begin{tabular}{|c|c|c|c|}
  \hline
  Category&Feature&Category&Feature \\
  \hline
  Time duration &  $t_u*(S-1)$  & Range & $B_1-B_2$\\
  Gradient &  $g_A=\max\{g_1,\ldots,g_S\}$ &  & $B_1-g_A$ \\
  &  $g_I=\min\{g_1,\ldots,g_S\}$ &  & $B_1-g_I$ \\
    &  $\overline{g}=\frac{1}{S}\sum_{i=1}^Sg_i$ &  & $B_2-g_A$ \\
   &  $Var=\sum_{i=1}^S(g_i-\overline{g})^2$ &  & $B_2-g_I$\\
  \hline
\end{tabular}
\end{table}

\begin{table}[htbp]
 \centering
\renewcommand{\arraystretch}{2.0}
\setlength\tabcolsep{4pt}
  \caption{Features of signal strength for action recognition. }\label{table:t3}
\begin{tabular}{|c|c|c|c|}
  \hline
  Feature&Definition&Feature&Definition \\
  \hline
  Average &  $\frac{1}{n}\sum_{i=1}^{n}x_i$  & Kurtosis & $\frac{\frac{1}{n}\sum_{i=1}^n(x_i-\overline{x})^4}{(\frac{1}{n}(x_i-\overline{x})^2)^2}-3$\\
  Range &  $x_{max}-x{min}$ & IQR & $Q_3-Q_1$ \\
  MAD&  $\frac{\sum_{i=1}{n}|x_i-\overline{x}|}{n}$ & Sum & $\sum_{i=1}^{n}x_i$ \\
  Variance  &  $\sum_{i=1}^{n}(x_i-\overline{x})^2$ & RMS & $\sqrt{\frac{\sum_{i=1}^{n}x_i^2}{N}}$ \\
  3rd C-Moment &  $\frac{1}{n}\sum_{i=1}^{n}x_i^3$ & Skewness & $\frac{\frac{1}{n}\sum_{i=1}^n(x_i-\overline{x})^3}{(\frac{1}{n}(x_i-\overline{x})^2)^{3/2}}$\\
  \hline
\end{tabular}
\end{table}

Suppose there are $\mathcal{N}$ \emph{quality classes} in the
classification. A quality class can be a driver for driver
identification or a body status for body status recognition. Also,
$\mathcal{N}$ is the number of drivers or body statuses. For a sample
(i.e., a 12-dimension vector), the NMLP computes an
$\mathcal{N}$-dimension normalized vector $V[1:\mathcal{N}]$, where
$\sum_{i=1}^\mathcal{N}V[i]=1$ and $V[i]$ is the probability that
the sample belongs to the $i$th class. In general, the $m$th class
is preferred when $V[m]\geq V[i]$ for all $1\leq i\leq \mathcal{N}$.
We use the NMLP to report the intermediate results such as $V$,
which plays an important role in the fusion process.

\textbf{Input of CNN}: The quality of actions can be characterized
in terms of the duration time and the speed and distance of
movement. We partition a signal fragment into ten segments and
extract the quality information from the three aspects. For each
segment, rather than the original gradient, a ten-dimension quality
vector is generated. Table~\ref{table:quality} summarizes the
quality vector, where $g_1,\ldots, g_S$ denote the gradient
sequence, $S$ the number of sampling points, $B_1$ the gradient at
the start point, and $B_2$ that at the end point. In total, the input
of CNN is a $10\times10$ matrix (or a 100-dimension vector).

\textbf{Action recognition}: We should first recognize the
action. The process is quite similar except for the input feature vector
and the number of classes, which are equal to that of all actions. To
generate the input vector, similarly, a fragment is divided into ten
segments and, for each segment, ten statistical features are
extracted. Table~\ref{table:t3} shows the definitions of features, where $x_i$ denote the signal strength at the $i$th
sampling point, $\overline{x}$ the average strength, and $n$ the
number of sampling points. 

\textbf{Feature selection}: Currently, there is no established theory to characterize the effect of different features or parameter choices on the action/quality recognition performance. It is of great significance to address such a fundamental problem. At this time, however, we have to choose the features according to the results presented in previous work and the characteristics of the concerned application.

First, to choose the features in Table~\ref{table:t3} for action recognition, we consider the series work  of Stephan Sigg et al as a reference~\cite{DBLP:journals/tmc/SiggSSJB14, DBLP:conf/huc/SiggSJ13, DBLP:conf/momm/SiggSBJW13}. These authors propose more than ten features of RSSI, such as the mean and variance, and investigate the discriminative capability of the features for action recognition. One of the findings is that, the effectiveness of features is tightly correlated with the signal propagation environment, and an adaptive policy is required in feature selection to achieve good performance.

Second, as shown before, the quality of actions is mainly captured by the gradient of signal strength variance. For example, when an action occurs suddenly and rapidly, the received signal strength should change sharply, resulting in a large gradient change. Therefore, we first obtain the gradient information at each moment, and then get the typical ``atomic'' statistics such as the mean, variance, and variation range of the gradient, as shown in Table~\ref{table:quality}.

For both quality recognition and action recognition, to avoid feature selection by hand and achieve high classification accuracy, we adopt a deep learning framework to automatically fuse the features by multi-layer nonlinear processing.

\subsection{Gradient-based signal boundary detection}

As the radio signal is sampled continuously, when multiple
actions occur sequentially, the start and end points of each
action must be located accurately. The signal is separated into many fragments, and each fragment corresponds to one action. As shown in Fig.~\ref{fig:fiveActivities}(b), there are usually three or more actions
in an activity to complete a driving task. To analyze the quality,
it is necessary to detect the signal boundary for each individual
action.

We propose to detect the signal boundary based on the gradient
changes of signal strength. As the signal strength begins to change
at the start point and becomes stable after the end of an action, it
is expected that the gradient could change sharply at the boundary
points. This is true for the actions related to the throttle (see
Fig.~\ref{fig:acceleratorClap}). For the actions related to the
clutch or the brake, there is another peak point in the received
signal sequence in addition to the boundary points. As a result, a turning
point can be detected by a sharp change in the gradient during the
execution of the action. Nevertheless, around the turning point, the
gradient always deviates from 0. In comparison, the gradient before
the start point or after the end point is close to 0.

\restylealgo{ruled}
\begin{algorithm}[htbp]
\small \linesnumbered \dontprintsemicolon

\KwData{Gradient sequence $GS[0:G-1]$}

\KwResult{$BP$, boundary point sequence}

y=L;

\Repeat{$y>G-L$}{
Compute the pre-average $a_r=\sum_{i=y-L}^{y-1}GS[i]/L$;

Compute the post-average $a_o=\sum_{i=y+1}^{y+L}GS[i]/L$;

\uIf{\{$abs(a_o)>\alpha*abs(a_r)$ and $abs(a_r)\leq \delta$ \}}{ Add
$y$ into $BP$ as a start point;
}

\uIf{\{$abs(a_r)>\alpha*abs(a_o)$ and $abs(a_o)\leq \delta$\}}{ Add
$y$ into $BP$ as an end point;
}

y=y+Step;
}

Prune the redundant boundary point in $BP$;

\caption{Computation of the boundary
points.}\label{Algorithm:boundaryPoints}
\end{algorithm}

The gradient-based boundary detection method is shown in Algorithm~\ref{Algorithm:boundaryPoints}. Basically, a sampling point is
regarded as the start of an action when (1) the average gradient
before the point approaches 0 and (2) the average gradient after the
point significantly deviates from 0. Alternately, a point is
regarded as the end of an action when (1) the average gradient
after the point approaches 0 and (2) the average gradient before the
point significantly deviates from 0.

An optimization framework is established to prune the
redundant points. The objection is to find the optimal number (e.g.,
$U$) of fragments and the intended sequence of fragments to satisfy
\begin{equation}\label{eq:boundary}
\begin{split}
\max\frac{1}{U}\sum_{u=1}^Up_{A(u)}
\end{split}
\end{equation}
where for the $u$th fragment, the recognized action is $A(u)$ with
probability $p_{A(u)}$. The advantage of (\ref{eq:boundary}) is that it is
simple, nonparametric and low in complexity. By incorporating more
constraints, such as the duration length, a more complex model can be
established, which can achieve higher precision.

\textbf{Parameter setting}: The idea of the proposed policy to detect the boundary is inspired by previous study on wireless communication~\cite{HalperinAW:mobicom08}. Unfortunately, the method does not have a theoretical analysis though it has been used widely. There are four parameters, two sliding parameters (i.e., \emph{L} and \emph{Step}) and two threshold parameters (i.e., $\alpha$ and $\delta$). In experiment, we empirically set \emph{L} = 5, \emph{Step} = 2, $\alpha$ = 5 and
$\delta$ = 0.5. Particularly, when the SNR is low, we set $\delta$ = 0.8.

Taking $\alpha$ as an example, we find that, even when there is no action, the received signal strength varies consistently and the range of variation (i.e., ratio of the maximum signal strength and the minimum one) can be as large as three. A similar conclusion was drawn in previous work~\cite{DBLP:conf/pimrc/El-KafrawyYE11}. Therefore, we set the threshold ($\alpha$) to five to achieve a good tradeoff between robustness and sensitivity. We also explore an adaptive policy to set the threshold. To determine the threshold used at time \emph{t}, we track the signal strength for a long time interval (approximately 1-2s) before time \emph{t}. We compute the ratio of the signal strength at each sampling point to the minimum one during the interval and choose \emph{x} as the threshold, where at least 90\% of the ratios are equal to or less than \emph{x}. The process is stopped when a start point is found and re-started when an ending point is detected. With the adaptive policy, the classification accuracy is close to the fixed setting used in our experiment. We plan to investigate adaptive policy improvements in the future. The processes to determine the other parameters are similar.

\subsection{Activity-based fusion}

Identification of the driver or body status based on a single action is vulnerable to noise or interference. To improve the accuracy of
quality recognition, WiQ adopts a fusion policy. In general,
multiple sensors or multiple classifiers are shown to increase the
recognition performance~\cite{DBLP:Polikar2006}.


%

We propose an activity-based fusion policy to exploit the temporal
diversity. The activity is chosen as the fusion unit for three
reasons. First, as all the actions in an activity are devoted to the
same driving task, the driving style should be stable. Second, as
the duration is not very long, it is expected that the wireless
channel does not vary drastically. Finally, as there are at least three or more
actions in an activity, it is sufficient to make a reliable decision
based on all of them together.

\begin{figure}
\begin{center}
\includegraphics*[scale=0.322]{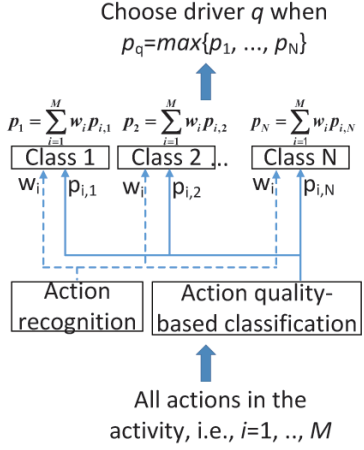}
\end{center}
\caption{Illustration of the fusion policy.} \label{fig:fusionbasic}
\end{figure}

A weighted majority voting rule is adopted. There are many available fusion rules, such as summation, majority voting,
Borda count and Bayesian fusion. Fig.~\ref{fig:fusionbasic} shows
the basic process of the fusion policy. Let $Q_1,\ldots,
Q_\mathcal{N}$ denote all the quality classes (e.g., drivers or body
statuses) and $A_1,\ldots, A_\mathcal{M}$ all the actions. Consider an
activity with $M$ actions denoted by $a_1,\ldots, a_M$. Without loss
of generality, suppose for each $a_i$, the action is classified as
$A_j$ with a probability of $w_i$. The role of $w_i$ is to capture the
effect of the channel condition (i.e., the better the channel is, the
higher $w_i$ is). In addition, letting $p(i,k)$ denote the probability that the
quality class of $a_i$ is $Q_k$ and $p_k$ be the probability that
the quality class of the activity is $Q_k$, we have
\begin{equation}\label{eq:fusion}
\begin{split}
p_k=\sum_{i=1}^Mw_i\times p(i,k).
\end{split}
\end{equation}

Finally, $Q_q$ is preferred as the quality class of the activity
when $p_q=\max\{p_1,\ldots, p_\mathcal{N}\}$.

\subsection{Discussions}

We now discuss some practical issues and possible extensions of WiQ.

\textbf{Efficiency}: Computational efficiency is known as one of the
major limitations of deep learning. As there is usually a large
number of parameters, the speed of a deep learning network is slow.
Thanks to the small network size, the efficiency of WiQ is very
high, e.g., only several microseconds are required to process the signals of an activity.

\textbf{Structure of activity}: In practice, the driving actions are
not completely random and instead usually follow a special order to complete
a driving task. It is expected that better performance will be achieved for
action recognition or signal boundary detection if the structure of the
activity is exploited.

\textbf{Online learning}: Currently, only after an entire driving
activity is completed can the signals be extracted for analysis. To
work online, there are several challenges such as noise reduction,
in-time boundary detection and exploitation of the history information to
facilitate the real-time quality recognition.

\textbf{Information fusion}: The fusion policy explored combines several intermediate classification results into a single
decision. Rather than combination, a boosting method can be adopted
to train a better single classifier gradually. Moreover, the
performance can be improved further by using numerous custom
classifiers dedicated to specific activity subsets. 

\section{PERFORMANCE EVALUATION}
\label{sec:performanceevaluation}

\begin{figure}
\begin{center}
\includegraphics*[scale=0.054]{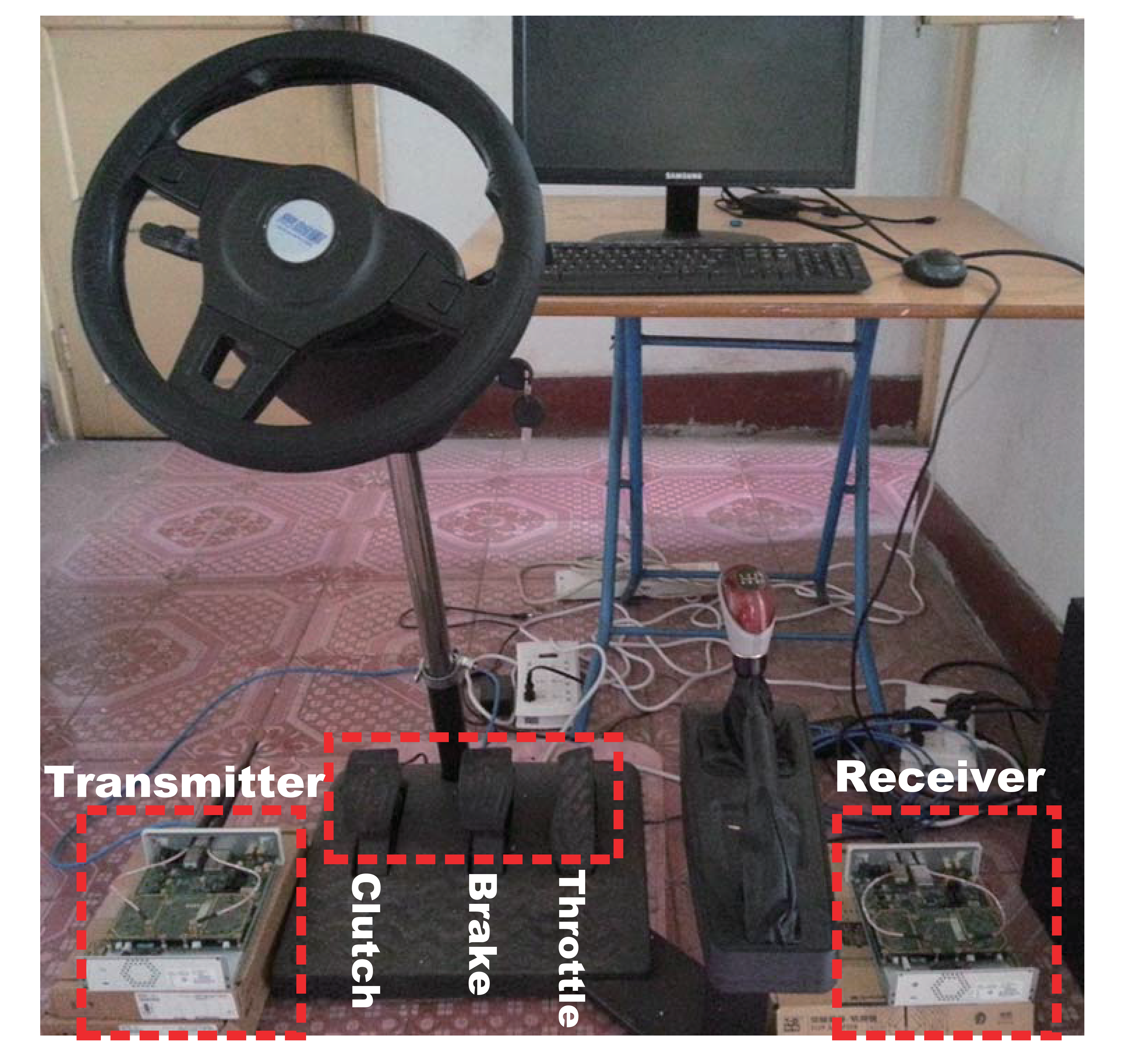}
\end{center}
\caption{Experimental setup with a driving emulator.}
\label{fig:expSetupDrive}
\end{figure}

We evaluate the performance by measurements in a testbed with a
driving emulator. Fig.~\ref{fig:expSetupDrive} shows the
experimental environment. The driving emulator includes three pedals: the clutch, brake and throttle. We use a software radio, the
Universal Software Radio Peripheral (USRP) N210~\cite{USRP:web}, as
the transmitter and receiver nodes. The signal is transmitted
uninterruptedly at 800MHz with 1Mbps data rate. The sampling rate is
200 samples per second at the receiver. 

The drivers are asked to perform all six activities shown in
Fig.~\ref{fig:fiveActivities}. The strategy is that (1) each driver repeats every activity 200 times regardless of the traffic conditions and (2) a driver drives on a given road (urban or high-speed road). If the number of activity execution is less than 200, the experiment is repeated until the number reaches 200 on the same road. The first strategy is adopted for the results presented in Section~\ref{sec:performanceevaluation} (A)-(C) and the second is adopted for Section~\ref{sec:performanceevaluation}-(D). For each action, there are approximately 400 samples. According to the average SNR, all the samples are equally divided into two categories, i.g., high-SNR (8-11 dB) and low-SNR (4-8 dB). The average SNR difference of the two categories is approximately 3.8dB.

The platform we used is a PC desktop with an 8-core Intel Core i7 CPU running at 2.4GHz and 8GB of memory. We do not use GPU to run the experiment. Unless otherwise specified, each data point is obtained by averaging the results from 10 runs.

\begin{figure*}
\begin{minipage}[b]{0.3\linewidth}
\begin{center}
\includegraphics*[scale=0.175]{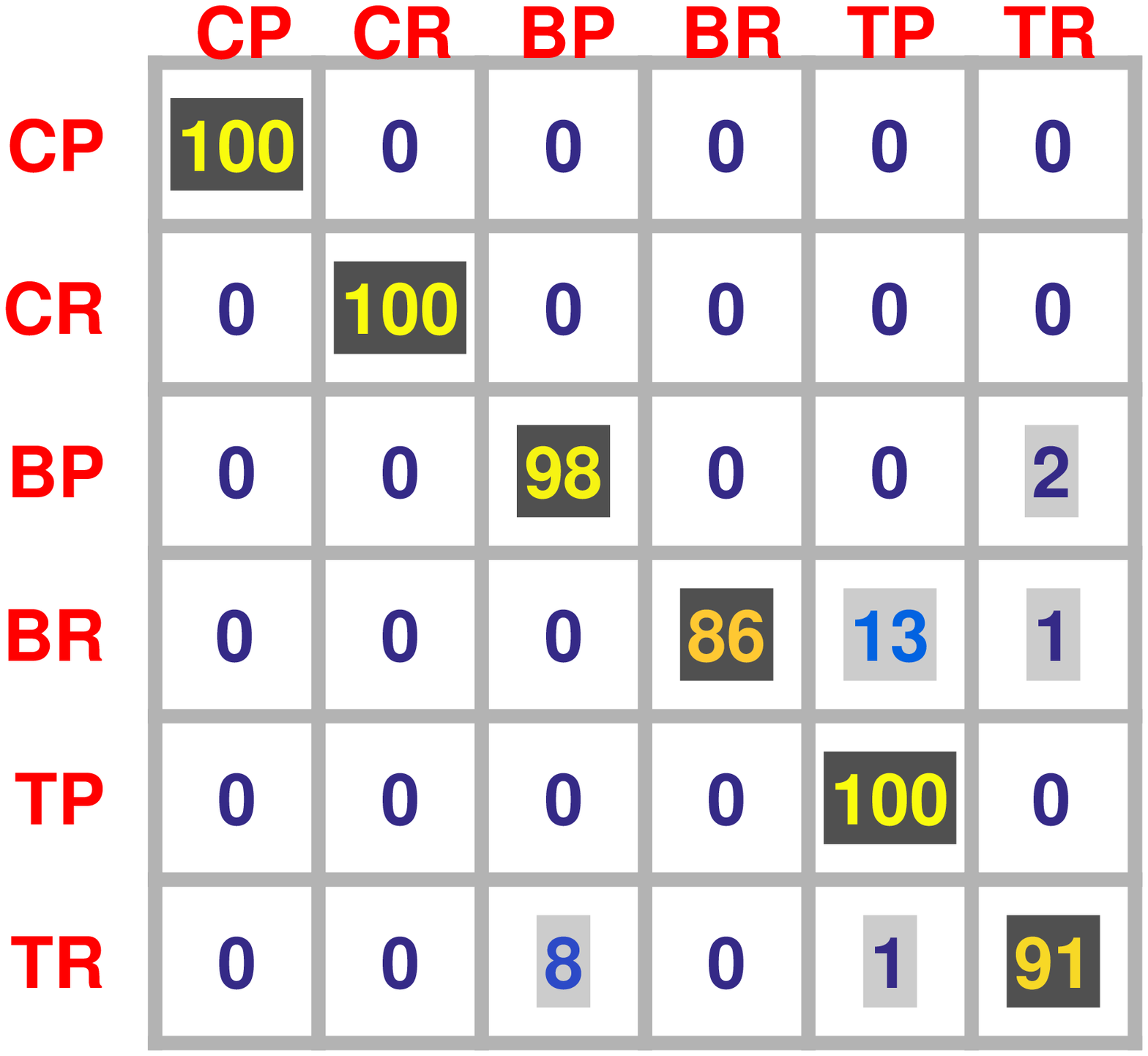}
\end{center}
\caption{Action recognition with high SNR and 10 training instances. }
\label{fig:action10}
\end{minipage}
\hspace{0.01\linewidth}
\begin{minipage}[b]{0.34\linewidth}
\begin{center}
\includegraphics*[scale=0.175]{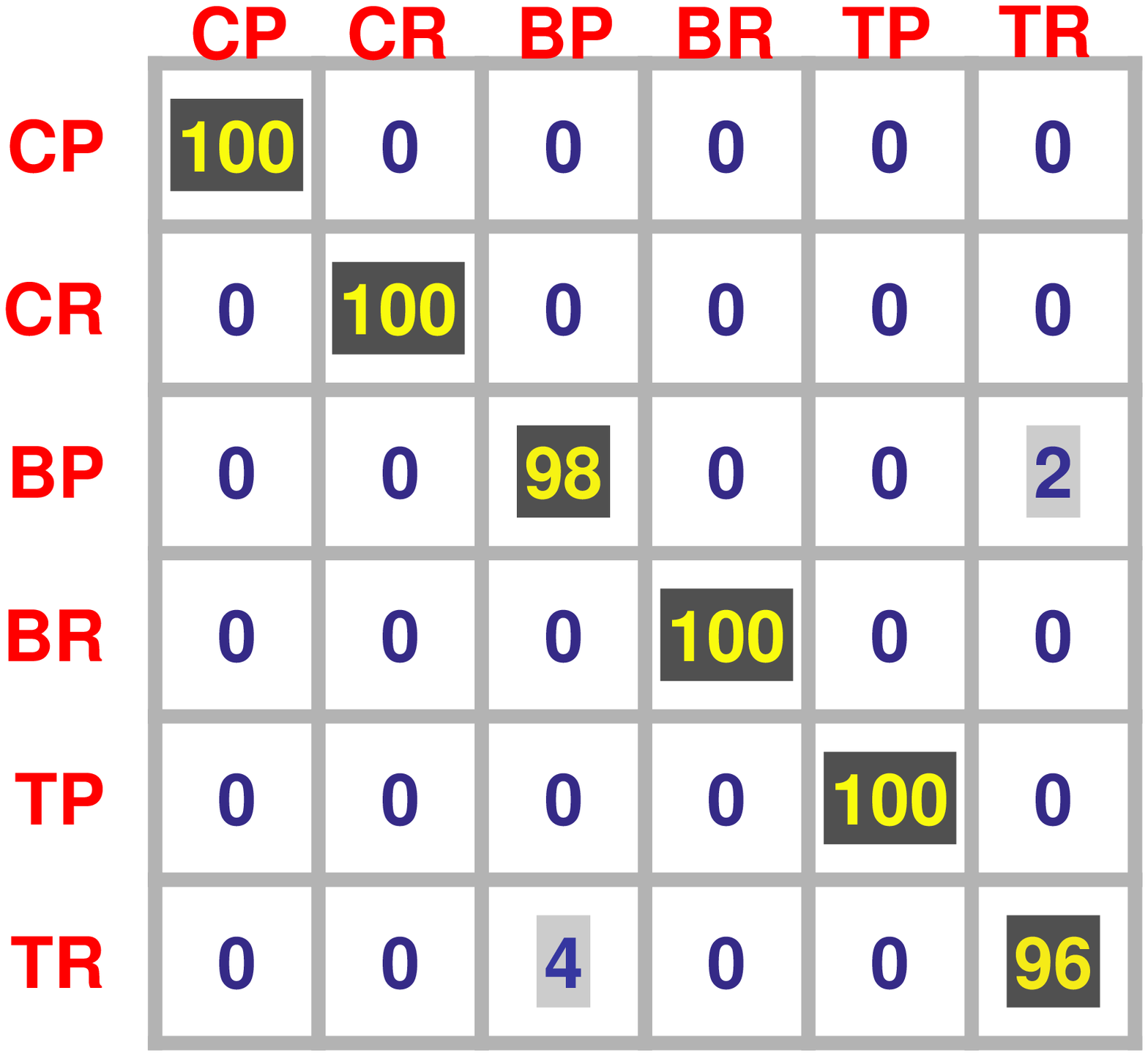}
\end{center}
\caption{Action recognition with high SNR and 100 training instances.
} \label{fig:action100}
\end{minipage}
\hspace{0.01\linewidth}
\begin{minipage}[b]{0.34\linewidth}
\begin{center}
\includegraphics*[scale=0.175]{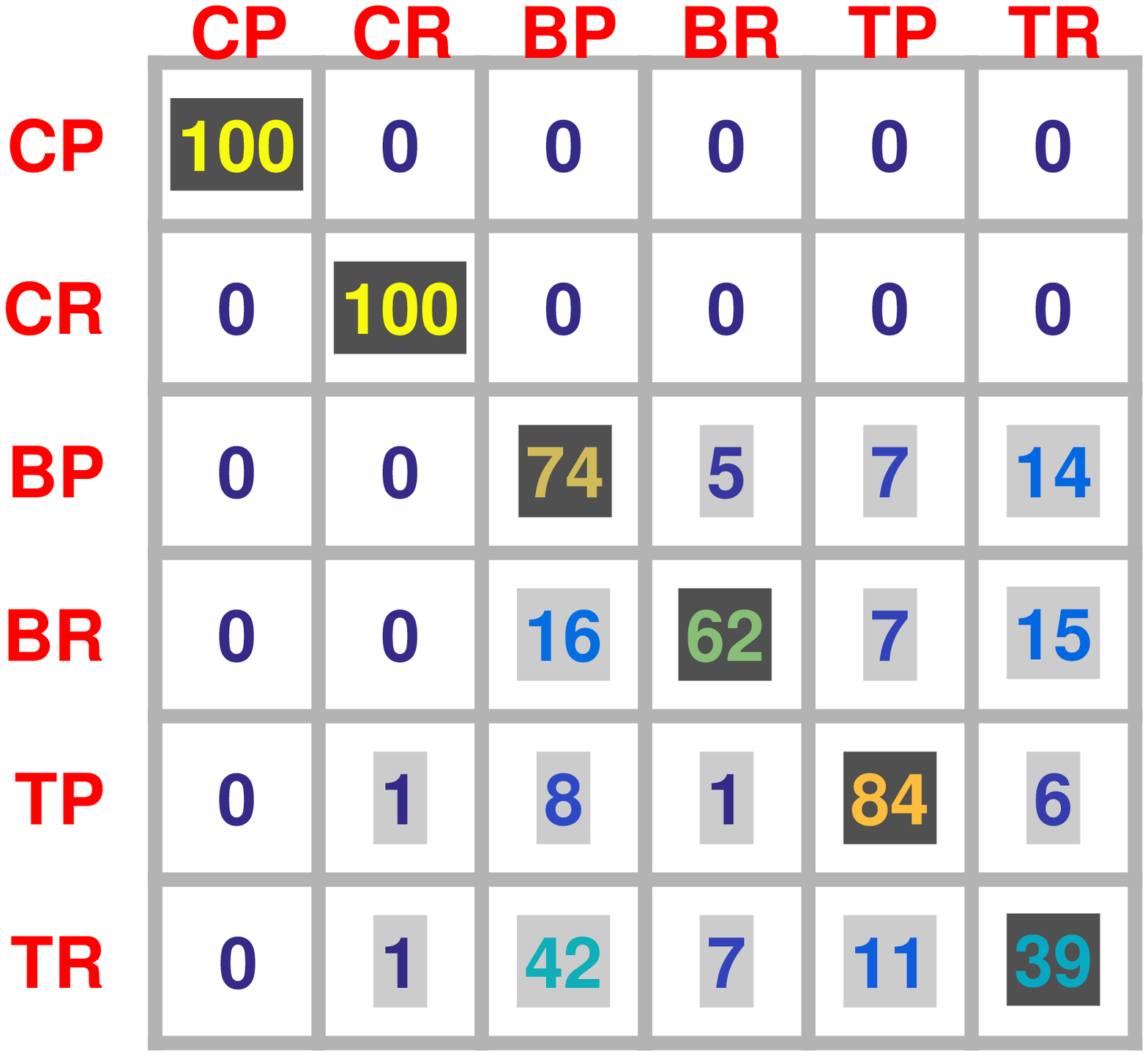}
\end{center}
\caption{Action recognition with low SNR and 100 training instances. }
\label{fig:actionLowSNR}
\end{minipage}
\end{figure*}

%
%

\subsection{Action recognition}

For each dataset in the high-SNR category, we choose 100 samples
randomly for training and the remaining for test. Fig.~\ref{fig:action10} and Fig.~\ref{fig:action100} show the results of recognition
accuracy. For example, the value (i.e., 13\%) at position (4, 5) is
the (error) probability that BR is recognized as TP. The recognition
accuracy is shown by the diagonal of the matrix. When the training
number of the CNN network is 10, the accuracy is at least 86\% and
on average 95\%. With more training (e.g., 100 times), the
performance becomes much better, i.e., the average accuracy
approaches 98\%. Nevertheless, the impact of noise or
interference is severe on the performance of action recognition. As
shown in Fig.~\ref{fig:actionLowSNR}, for the low-SNR category, the
accuracy is as low as 39\% and  on average 65\%.

\begin{figure}
\begin{center}
\includegraphics*[scale=0.262]{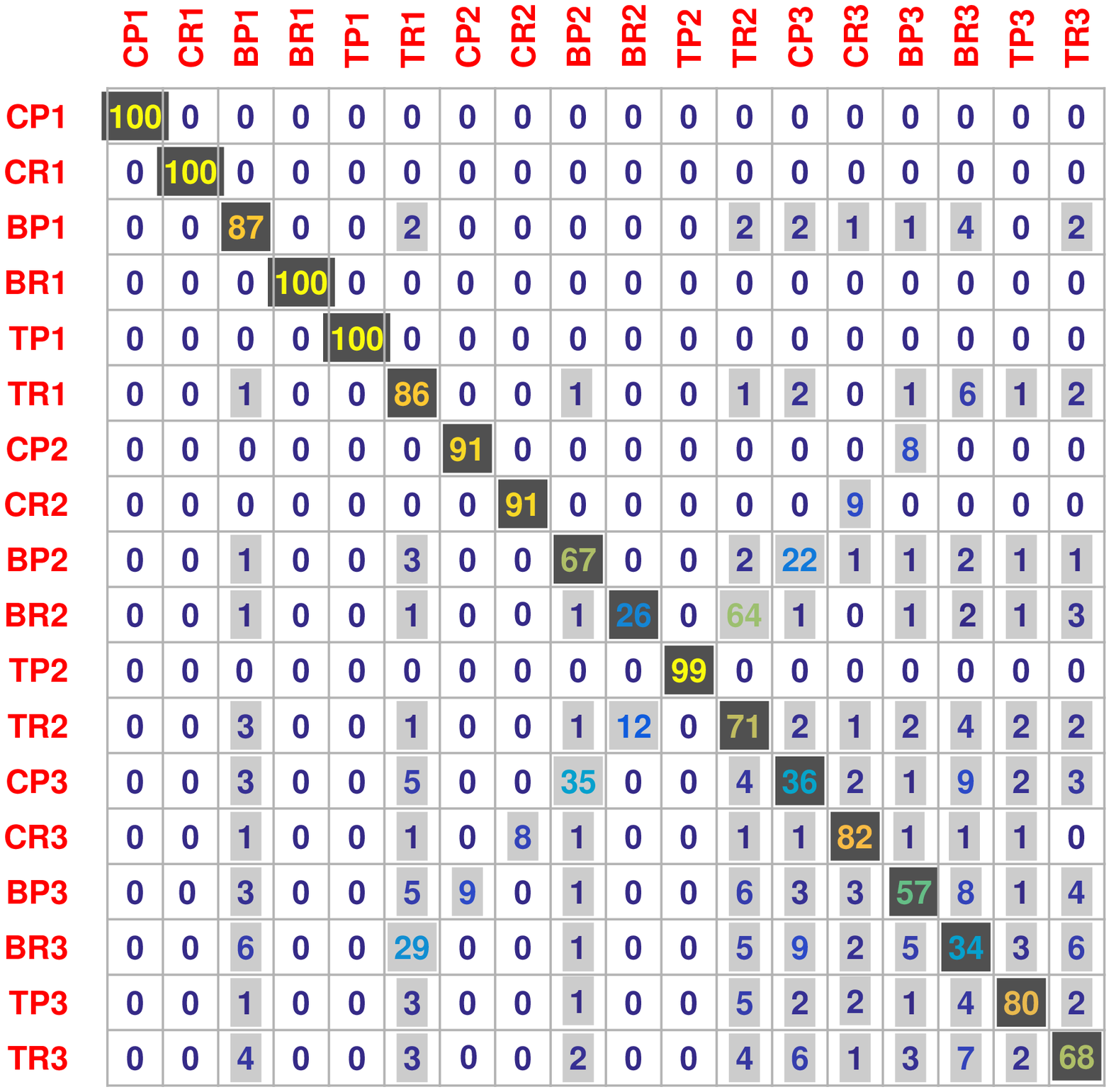}
\end{center}
\caption{Action recognition with multiple drivers, high SNR and 1000
training instances.} \label{fig:3Drivers18Class}
\end{figure}

More experiments are conducted with three drivers. Together with the
six actions, we have a total of 18 classes. For each class, from the
high-SNR samples, we randomly select 100 of them for training and
the remaining for test. Fig.~\ref{fig:3Drivers18Class} shows the
results when the training number is 1000. As the number of class is
much larger, the accuracy decreases drastically, which can be as low as
26\% and approximately 60\% on average.

At the same time, there is a large number of cross-driver errors,
i.e., the action of a driver is recognized as that of the other one.
For example, the error probability between CP3 and BP2 is 35\% (CP3
to BP2) and 22\% (BP2 to CP3). As a result, there would be much more
mistakes if we try to identify the driver based on the action alone.

\subsection{Capability of quality recognition}

We now investigate the capability of the quality recognition in an
intuitive manner. For simplicity, we consider two dimensions of the
quality, i.e., average gradient and duration.

\begin{figure}
%
\begin{center}
\includegraphics*[scale=0.92]{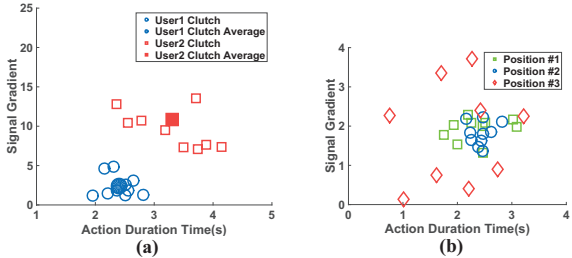}
\end{center}
\caption{The quality distribution for clutch-pressing with (a)
different drivers or (b) different receiver positions.}
\label{fig:rssDiffPeople}
\end{figure}

First, we investigate the ability to distinguish the drivers. Fig.~\ref{fig:rssDiffPeople} (a) shows the quality distribution for
clutch-pressing with different drivers. The points can be
clustered into two categories. Meanwhile, the difference between
different clusters is quite significant. That is, the driving style is stable
for the same person but distinct for different drivers.

\begin{figure}
\begin{center}
\includegraphics*[scale=0.182]{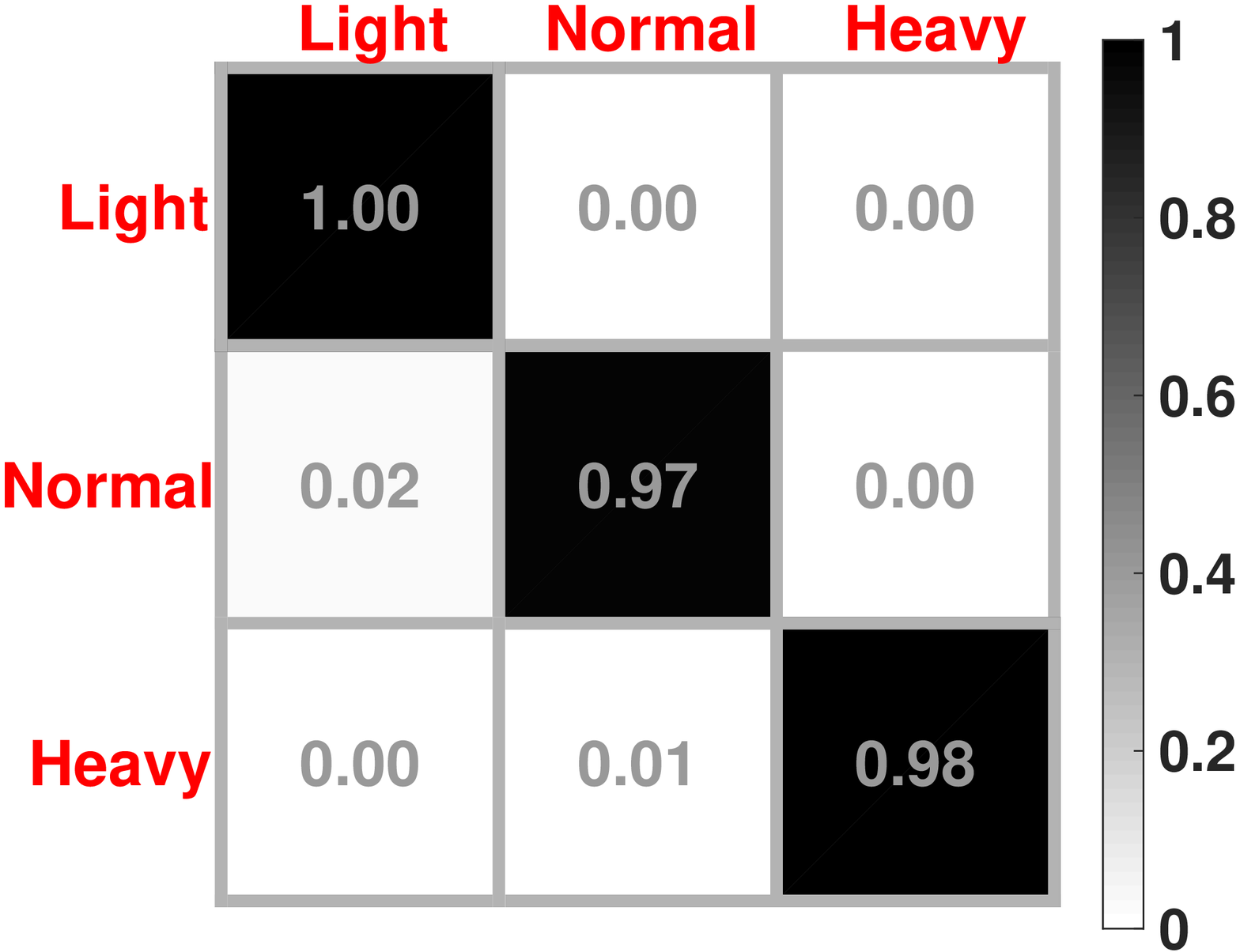}
\end{center}
\caption{Body status recognition based on the quality of action with high SNR.}
\label{fig:qualityBodystatus}
\end{figure}

\begin{figure*}
\begin{minipage}[b]{0.3\linewidth}
\begin{center}
\includegraphics*[scale=0.185]{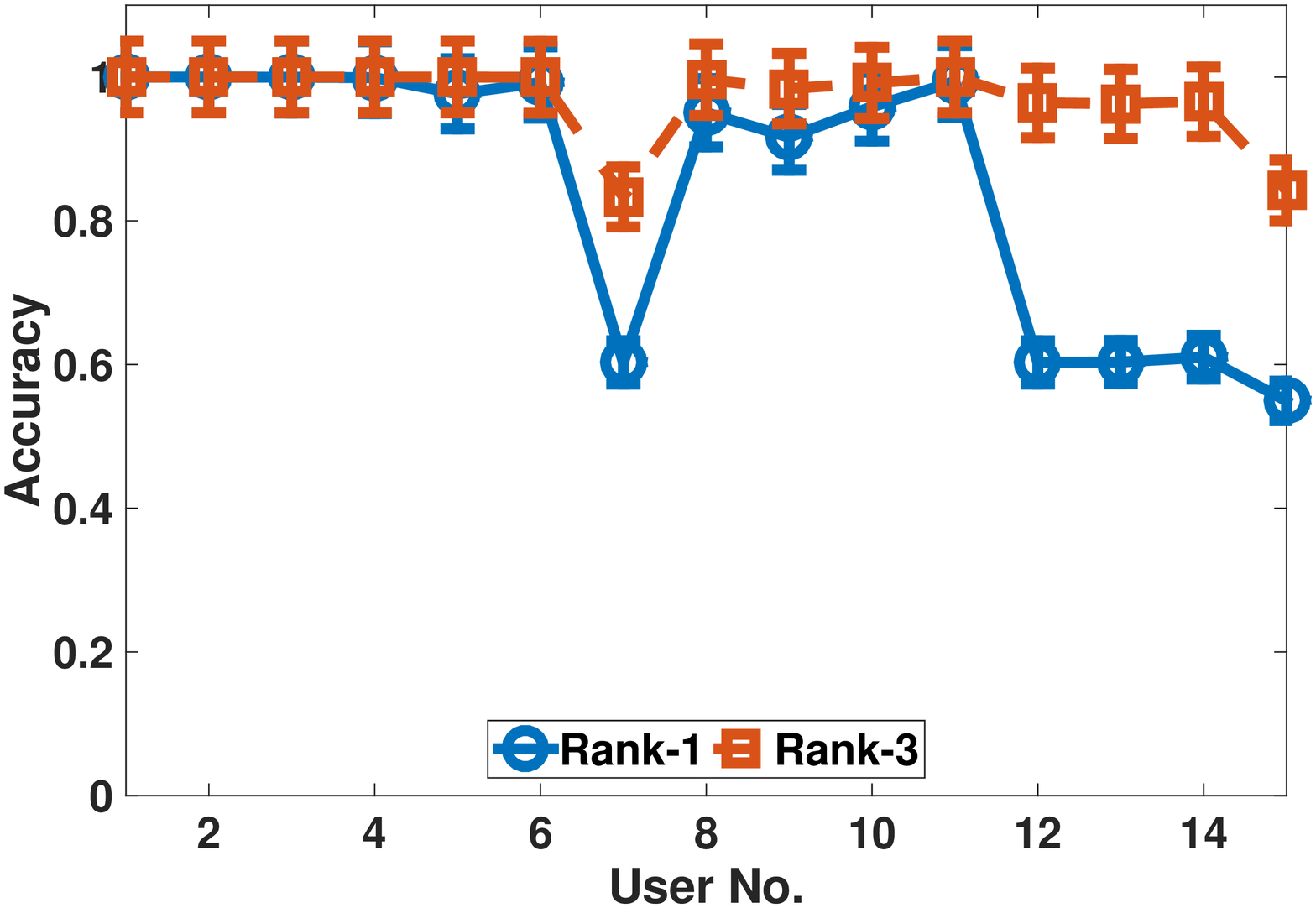}
\end{center}
\caption{The accuracy of driver identification without fusion in the
high-SNR category. } \label{fig:actionDriverIdent}
\end{minipage}
\hspace{0.01\linewidth}
\begin{minipage}[b]{0.34\linewidth}
\begin{center}
\includegraphics*[scale=0.185]{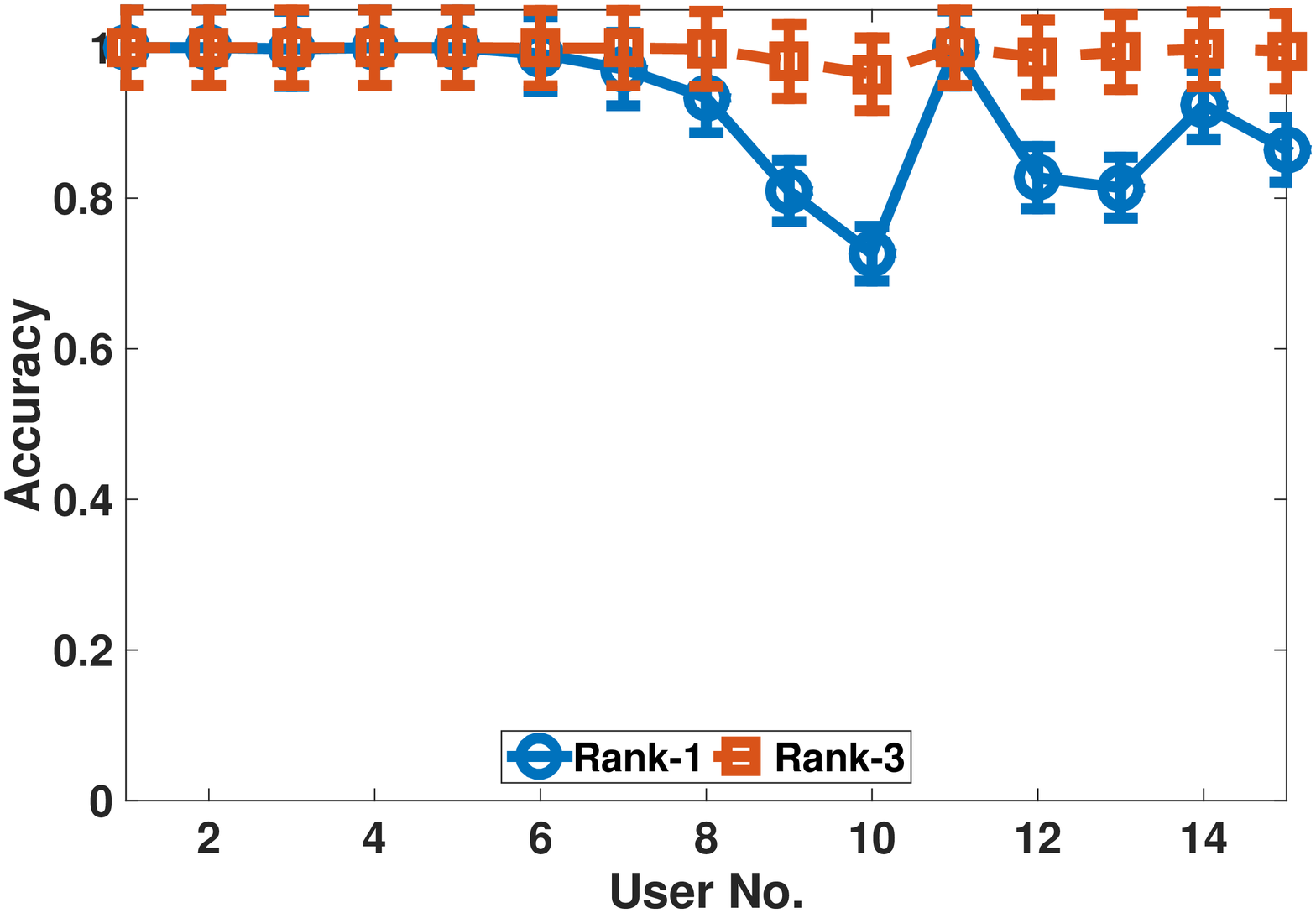}
\end{center}
\caption{The accuracy of driver identification with activity-based
fusion in the high-SNR category. }
\label{fig:activityDriverIdent-High}
\end{minipage}
\hspace{0.01\linewidth}
\begin{minipage}[b]{0.34\linewidth}
\begin{center}
\includegraphics*[scale=0.185]{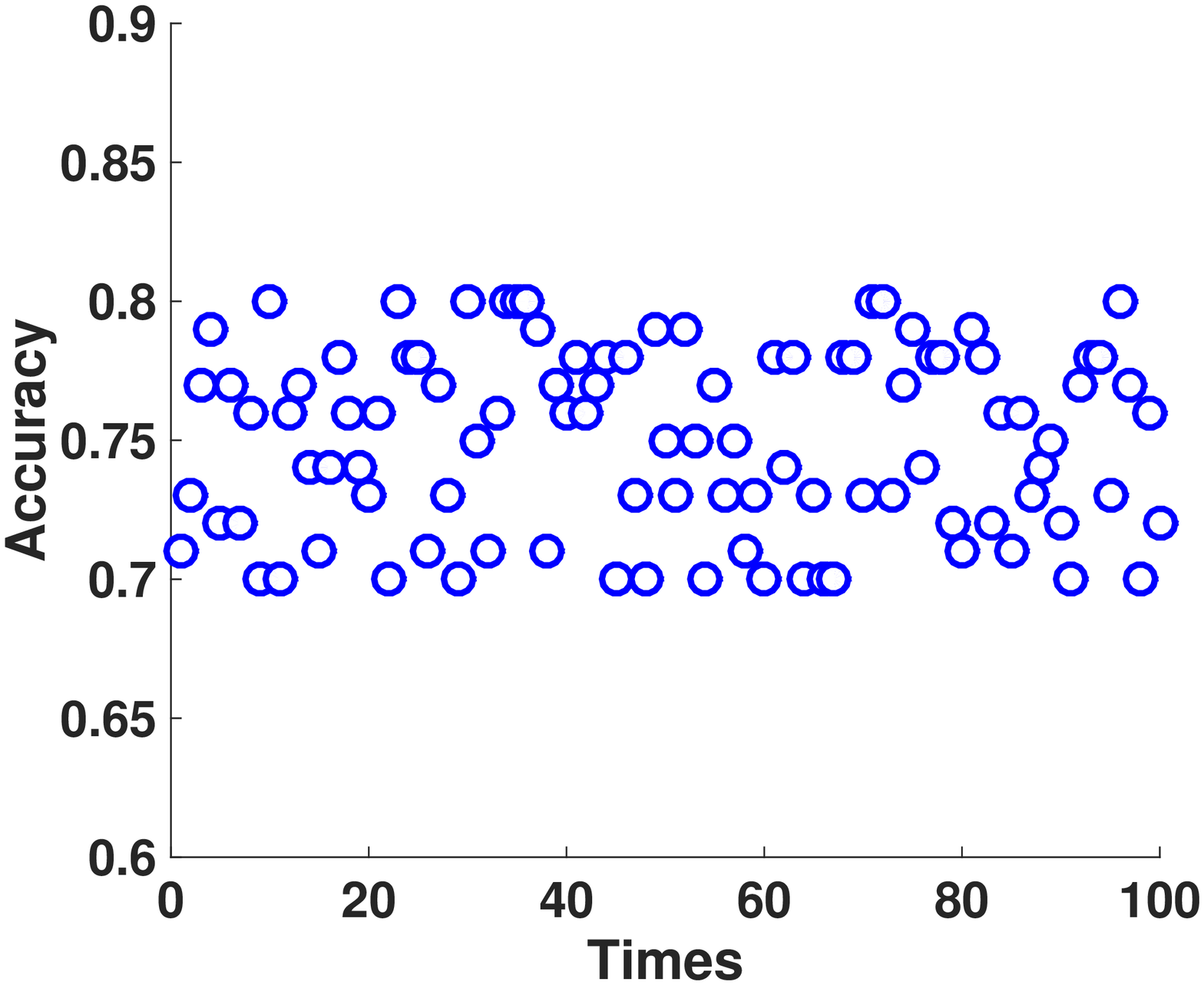}
\end{center}
\caption{The accuracy of driver identification with activity-based
fusion for all samples. } \label{fig:activityDriverIdent-Low}
\end{minipage}
\end{figure*}

Second, the sensitivity to the receiver position is investigated.
Fig.~\ref{fig:rssDiffPeople} (b) shows the quality distribution for
CP with three receiver positions. Similarly, the points can be
categorized into three groups, indicating the dependence of the
quality on the receiver position. In wireless communications, even
when the receiver position is changed slightly, the signal
propagation characteristics can vary drastically. In practice, when
the node position is changed, the convolutional neural network
should be re-trained. In the following, the experiments are performed with the same receiver position, i.e., the position \#1 in Fig.~\ref{fig:rssDiffPeople} (b).

\subsection{Application with quality recognition}

We investigate the performance of qualitative action recognition. The
training number of CNN is 100 by default.

Consider body status recognition first. As it is not easy to carry out experiments to detect the fatigue status, our focus turns to the detection of attention. WiQ tries to distinguish the three body statuses: (1)
normal, the normal state; (2) light distraction, i.e., driving a car
while reading a slowly changing text (5 words per second); and (3)
heavy distraction, i.e., driving a car while reading a rapidly
changing text (15 words per second). We use the 200 high-SNR samples
in the experiment: 100 samples are selected randomly to train the
neural network and the remaining samples are utilized for testing. As
shown in Fig.~\ref{fig:qualityBodystatus}, 
the average accuracy is as high as 97\%. The results indicate that the
quality information is very useful in distinguishing the body
condition of a driver.

Now consider driver identification. When the number of drivers is large,
it is much more challenging than the recognition of the body status.
There are 15 drivers in the experiments, among which three have five
years or more of driving experience, five are novices, and the rest
have 1$-$3 years of experience.

First, the drivers are identified based on the quality of their
individual actions. There are 15 driver classes and
Rank-\emph{k} means that, for a test sample, all classes are ranked
according to the probability computed by WiQ in descending order (the correct class belongs to the set of the first \emph{k} classes).
Fig.~\ref{fig:actionDriverIdent} shows the Rank-1 and Rank-3
recognition accuracy. The Rank-1 accuracy is at least 56\% and on average 78\%. In comparison with the results shown in Fig.~\ref{fig:3Drivers18Class}, by using the quality information, the
ability to identify the drivers is improved significantly. The
Rank-3 accuracy was at least 82\% and on average 95\%.

Second, the performance can be improved further by the
activity-based fusion policy. For each activity, there are 200
samples. We partition them equally into the high-SNR and low-SNR
categories. Afterwards, we select 60 high-SNR samples randomly for
training and the remaining for testing. Fig.~\ref{fig:activityDriverIdent-High} shows the results. The Rank-1
accuracy is always higher than 72\% and on average 88\%. Additionally, the
Rank-3 accuracy approaches 97\%. In other words, high identification
precision can be achieved by WiQ when the SNR is high.

For the low SNR scenario, we choose the set of test samples
randomly from the 100 low-SNR samples. The experiment was repeated
for approximately 100 times. Fig.~\ref{fig:activityDriverIdent-Low} plots
the average Rank-1 accuracy. Though the accuracy is lower than that
in the high-SNR category, promising performance is achieved with
the help of the fusion strategy, i.e., the accuracy is as high as 80\% and
on average 75\%.

In summary, WiQ can recognize the action accurately and discriminate
among different body statuses (or drivers) based on the driving
quality. For action recognition, the accuracy is as high as 95\%
when the SNR is high. In addition, the accuracy of body status recognition
is as high as 97\%. For driver identification, the average Rank-1
accuracy is 88\% with high SNR and 75\% when the SNR is low.

\subsection{Comparative study}

We present the results of the comparative study. We first compare our method with other machine learning methods. Then, we present the sensitivity results of quality recognition on the gradient features. Finally, we discuss the driver category recognition (i.e., finding the category for a given driver) under various traffic conditions (urban vs. high-speed road). In general, there are three driver categories, ``Experienced'' ($>$3 years of driving experience), ``Less experienced'' (1-3 years of driving experience) and ``Novice'' ($<$1 year of driving experience). In comparison with driver identification, driver category recognition is a similar but easier task. The category information is useful in practice. For example, the driving assistant system can give more operable driving instructions to novice drivers and more alert information to experienced drivers.

\begin{table}[htbp]
 \centering
\renewcommand{\arraystretch}{2.0}
\setlength\tabcolsep{4pt}
  \caption{Average error rate of action recognition of CNN, SVM and \emph{k}NN with high SNR and different numbers of iterations. $k$NN does not need iteration, and the result is shown when \emph{k}=3.}\label{table:comparasionSVMkNNhigh}
\begin{tabular}{|c|cccccc|}
  \hline
  Iteration Number&5&10&15&20&50&100 \\
  \hline
  CNN&0.36&0.06&0.05&0.04&0.03&0.01\\
  SVM&0.52&0.41&0.34&0.30&0.28&0.27\\
  \emph{k}NN&&&0.35&&&\\
  \hline
\end{tabular}
\end{table}

\begin{table}[htbp]
 \centering
\renewcommand{\arraystretch}{2.0}
\setlength\tabcolsep{4pt}
  \caption{Average error rate of action recognition of CNN, SVM and \emph{k}NN with low SNR and different numbers of iterations. The result of \emph{k}NN is shown when \emph{k}=3.}\label{table:comparasionSVMkNNlow}
\begin{tabular}{|c|cccccc|}
  \hline
  Number of iterations&5&10&15&20&50&100 \\
  \hline
  CNN&0.36&0.06&0.05&0.04&0.03&0.01\\
  SVM&0.52&0.41&0.34&0.30&0.28&0.27\\
  \emph{k}NN&&&0.44&&&\\
  \hline
\end{tabular}
\end{table}

To demonstrate the effectiveness of the deep convolutional neural network (CNN), we choose \emph{k}-nearest neighbor (\emph{k}NN) and support vector machine (SVM)~\cite{DBLP:journals/tist/ChangL11} for comparison. For \emph{k}NN, we choose \emph{k}=3 as it achieves the best performance in the experiment. Table~\ref{table:comparasionSVMkNNhigh} shows the average error rate of action recognition with different numbers of iterations for CNN and SVM with high SNR. Table~\ref{table:comparasionSVMkNNlow} shows the results with low SNR. First, with a large number of iterations, the precision of CNN is very high, i.e., the error rate is only 1\% with high SNR. Even with low SNR, the average precision is still larger than 70\%. Second, CNN outperforms SVM significantly and consistently. Though the error rate of SVM decreases with an increase of the number of iterations, it is never lower than 27\% with high SNR or 38\% with low SNR. The performance of \emph{k}NN does not depend on the number of iterations and is consistently worse than that of SVM and CNN.

\begin{table}[htbp]
 \centering
\renewcommand{\arraystretch}{2.0}
\setlength\tabcolsep{4pt}
  \caption{Average error rate of body status recognition with SVM and different sets of gradient features. The best performance is achieved with ``G(3)+R(A)''.}\label{table:differentFeaturesSVM}
\begin{tabular}{|c|ccccccc|}
  \hline
  SNR&G(3)&G(A)&R(3)&R(A)&G(3)+R(A)&G(A)+R(A)&All \\
  \hline
  Low&0.49&0.36&0.50&0.42&0.41&0.30&0.31\\
  HIgh&0.28&0.21&0.34&0.26&0.27&0.14&0.17\\
  \hline
\end{tabular}
\end{table}

We investigate the sensitivity of quality recognition on the gradient features. As the process of feature fusion in WiQ is automatic, we choose SVM to conduct the experiment. Table VI shows the average error rate of body status recognition by  SVM  with 100 iterations. The features are selected from Table~\ref{table:t1}, where ``G(3)'' refers to
\{$g_A$, $g_I$, $\bar{g}$\}, ``R(3)'' to \{$B_1-B_2$, $B_1-g_A$, $B_1-g_I$\}, ``G(A)'' to
\{$g_A$, $g_I$, $\bar{g}$, $Var$\}, and ``R(A)'' to the five range features on the right
of Table~\ref{table:t1}. In general, a lower error rate can be achieved with
more features, except that the result of ``G(A)+R(A)'' is better than that when all features are used. Comparing the results in ``G(3)'' with those in ``G(A),'' one can observe that the second-order metric (i.e., the variance of the gradient) is quite effective for reducing the error rate, e.g., by 14\% with low SNR and 5\% with high SNR. It is generally insufficient to use the first-order statistics alone in action quality recognition, i.e., the error rate is as high as 27\% with high SNR in ``G(3)+R(A)'' where all the first-order statistics (except time duration) are used.

\begin{table}[htbp]
 \centering
\renewcommand{\arraystretch}{2.0}
\setlength\tabcolsep{4pt}
  \caption{Driver category recognition with high SNR on the urban road. Higher precision is achieved for novices as they have consistent non-optimal driving behaviors.}\label{table:categoryurban}
\begin{tabular}{|c|ccc|}
  \hline
  &Exp.&Less Exp.&Novice \\
  \hline
  Exp.&0.85&0.09&0.06\\
  Less Exp.&0.10&0.87&0.03\\
  Novice&0.03&0.07&0.9\\
  \hline
\end{tabular}
\end{table}

\begin{table}[htbp]
 \centering
\renewcommand{\arraystretch}{2.0}
\setlength\tabcolsep{4pt}
  \caption{Driver category recognition with high SNR on the high-speed road. Novices can be recognized accurately.}\label{table:categoryhighspeed}
\begin{tabular}{|c|ccc|}
  \hline
  &Exp.&Less Exp.&Novice \\
  \hline
  Exp.&0.92&0.06&0.02\\
  Less Exp.&0.05&0.89&0.06\\
  Novice&0.01&0.01&0.98\\
  \hline
\end{tabular}
\end{table}

Finally, Table~\ref{table:categoryurban} and Table~\ref{table:categoryhighspeed} show the results of driver category recognition on the urban and high-speed roads, respectively. The results are obtained with high SNR. One can see that the accuracy of quality recognition is lower in the urban environment. A possible reason is that a driver should react differently to distinct traffic condition on the urban road, resulting in difficulty in quality assessment. Moreover, the results of novice are relatively better. This is because a novice driver cannot adapt well to different traffic conditions, resulting in unified (but not optimal) reaction behavior.

In summary, the comparative study indicates first, that the deep neural network method outperforms \emph{k}NN and SVM consistently; second, that second-order statistics, such as variance, are critical for achieving high performance of quality recognition; and third, that it is more challenging to recognize driving quality under complex traffic conditions (e.g., urban roads).

\section{CONCLUSIONS AND FUTURE WORK}
\label{sec:Conclusions}

We take the driving system as an example of human-machine system and study the fine-grained recognition of driving behaviors. Although action recognition has been studied extensively, the quality of actions is less understood. We propose WiQ for qualitative action recognition by using narrowband radio signals. It has three key components, deep neural network-based learning, gradient-based signal boundary detection, and activity-based fusion. Promising performance is achieved for the challenging applications, e.g., the accuracy is on average 88\% for identification among 15 drivers. Currently, the experiments are performed with a driving emulator. In the future, we plan to further optimize the learning framework and evaluate the performance of the proposed method in a real environment.


\section*{Acknowledgment}

This work has been supported by the NSF of China (No. 61572512, U1435219 and 61472434). The authors sincerely appreciate  the reviewers and editors for their constructive comments.


\bibliographystyle{IEEEtran}
\bibliography{sic}

\begin{thebibliography}{10}
\providecommand{\url}[1]{#1}
\csname url@samestyle\endcsname
\providecommand{\newblock}{\relax}
\providecommand{\bibinfo}[2]{#2}
\providecommand{\BIBentrySTDinterwordspacing}{\spaceskip=0pt\relax}
\providecommand{\BIBentryALTinterwordstretchfactor}{4}
\providecommand{\BIBentryALTinterwordspacing}{\spaceskip=\fontdimen2\font plus
\BIBentryALTinterwordstretchfactor\fontdimen3\font minus
  \fontdimen4\font\relax}
\providecommand{\BIBforeignlanguage}[2]{{%
\expandafter\ifx\csname l@#1\endcsname\relax
\typeout{** WARNING: IEEEtran.bst: No hyphenation pattern has been}%
\typeout{** loaded for the language `#1'. Using the pattern for}%
\typeout{** the default language instead.}%
\else
\language=\csname l@#1\endcsname
\fi
#2}}
\providecommand{\BIBdecl}{\relax}
\BIBdecl

\bibitem{DBLP:conf/infocom/AbdelnasserYH15}
H.~Abdelnasser, M.~Youssef, and K.~A. Harras, ``Wi{G}est: A ubiquitous
  {WiFi}-based gesture recognition system,'' \emph{\emph{in} Proc. {IEEE}
  {INFOCOM}'15}, pp. 75--86, 2015.

\bibitem{DBLP:journals/tmc/GuoCHYZW16}
B.~Guo, H.~Chen, Q.~Han, Z.~Yu, D.~Zhang, and Y.~Wang, ``Worker-contributed
  data utility measurement for visual crowdsensing systems,'' \emph{{IEEE}
  Trans. Mob. Comput.}, vol.~PP, no.~99, pp. 1--1, 2016.

\bibitem{DBLP:journals/thms/YuXYG15}
Z.~Yu, H.~Xu, Z.~Yang, and B.~Guo, ``Personalized travel package with
  multi-point-of-interest recommendation based on crowdsourced user
  footprints,'' \emph{{IEEE} Trans. on Human-Machine Systems}, vol.~46, no.~1,
  pp. 151--158, 2016.

\bibitem{DBLP:journals/corr/GuoLWYH16}
B.~Guo, Y.~Liu, W.~Wu, Z.~Yu, and Q.~Han, ``Activecrowd: {A} framework for
  optimized multi-task allocation in mobile crowdsensing systems,''
  \emph{{IEEE} Trans. on Human-Machine Systems}.

\bibitem{DBLP:conf/aughuman/VellosoBGUF13}
E.~Velloso, A.~Bulling, H.~Gellersen, W.~Ugulino, and H.~Fuks, ``Qualitative
  activity recognition of weight lifting exercises,'' \emph{\emph{in} Proc. ACM
  AH'13}, pp. 116--123, 2013.

\bibitem{fatiguereview12}
{Caterpillar}, ``Operator fatigue detection technology review,''
  \emph{Caterpillar Global Mining}, pp. 1--58, 2012.

\bibitem{DBLP:journals/tvt/JiZL04}
Q.~Ji, Z.~Zhu, and P.~Lan, ``Real-time nonintrusive monitoring and prediction
  of driver fatigue,'' \emph{{IEEE} T. Vehicular Technology}, vol.~53, no.~4,
  pp. 1052--1068, 2004.

\bibitem{DBLP:conf/chi/VellosoBG13}
E.~Velloso, A.~Bulling, and H.~Gellersen, ``Motionma: motion modelling and
  analysis by demonstration,'' \emph{\emph{in} Proc. ACM {CHI} '13}, pp.
  1309--1318, 2013.

\bibitem{USRP:web}
USRP, ``Ettus research,'' \emph{http://www.ettus.com}, 2010.

\bibitem{DBLP:journals/corr/Wang0T15}
L.~Wang, Y.~Qiao, and X.~Tang, ``Action recognition with trajectory-pooled
  deep-convolutional descriptors,'' \emph{\emph{in} Proc. {IEEE} {CVPR}'15},
  pp. 4305--4314, 2015.

\bibitem{DBLP:conf/chi/CohnMPT12}
G.~Cohn, D.~Morris, S.~Patel, and D.~S. Tan, ``Humantenna: using the body as an
  antenna for real-time whole-body interaction,'' \emph{\emph{in} Proc. ACM
  {CHI}'12}, pp. 1901--1910, 2012.

\bibitem{DBLP:conf/chi/GuptaMPT12}
S.~Gupta, D.~Morris, S.~Patel, and D.~S. Tan, ``Soundwave: using the doppler
  effect to sense gestures,'' \emph{\emph{in} Proc. ACM {CHI} '12}, pp.
  1911--1914, 2012.

\bibitem{DBLP:conf/nsdi/AdibKK15}
F.~Adib, Z.~Kabelac, and D.~Katabi, ``Multi-person localization via rf body
  reflections,'' \emph{\emph{in} Proc. {USENIX} {NSDI}'15}, pp. 279--292, 2015.

\bibitem{DBLP:conf/nsdi/JoshiBKK15}
K.~Joshi, D.~Bharadia, M.~Kotaru, and S.~Katti, ``Wideo: Fine-grained
  device-free motion tracing,'' \emph{\emph{in} Proc. {USENIX} {NSDI}'15}, pp.
  189--202, 2015.

\bibitem{DBLP:conf/mobicom/WangLCG0L14}
Y.~Wang, J.~Liu, Y.~Chen, M.~Gruteser, J.~Yang, and H.~Liu, ``E-eyes:
  device-free location-oriented activity identification using fine-grained wifi
  signatures,'' \emph{\emph{in} Proc. ACM MOBICOM'14}, pp. 617--628, 2014.

\bibitem{DBLP:conf/sigcomm/AdibK13}
F.~Adib and D.~Katabi, ``See through walls with wifi!'' \emph{\emph{in} Proc.
  {ACM} {SIGCOMM}'13}, pp. 75--86, 2013.

\bibitem{DBLP:conf/mobicom/PuGGP13}
Q.~Pu, S.~Gupta, S.~Gollakota, and S.~Patel, ``Whole-home gesture recognition
  using wireless signals,'' \emph{\emph{in} Proc. ACM MOBICOM'13}, pp. 27--38,
  2013.

\bibitem{DBLP:conf/nsdi/AdibKKM14}
F.~Adib, Z.~Kabelac, D.~Katabi, and R.~C. Miller, ``3{D} tracking via body
  radio reflections,'' \emph{\emph{in} Proc. {USENIX} {NSDI}'14}, pp. 317--329,
  2014.

\bibitem{DBLP:conf/huc/MelgarejoZRC14}
P.~Melgarejo, X.~Zhang, P.~Ramanathan, and D.~Chu, ``Leveraging directional
  antenna capabilities for fine-grained gesture recognition,'' \emph{\emph{in}
  Proc. ACM UbiComp'14}, pp. 541--551, 2014.

\bibitem{DBLP:journals/csur/YangZL13}
Z.~Yang, Z.~Zhou, and Y.~Liu, ``From {RSSI} to {CSI:} indoor localization via
  channel response,'' \emph{{ACM} Comput. Surv.}, vol.~46, no.~2, p.~25, 2013.

\bibitem{DBLP:conf/infocom/XiZLZTLJ14}
W.~Xi, J.~Zhao, X.~Li, K.~Zhao, S.~Tang, X.~Liu, and Z.~Jiang, ``Electronic
  frog eye: Counting crowd using wifi,'' \emph{\emph{in} Proc. IEEE
  INFOCOM'14}, pp. 361--369, 2014.

\bibitem{DBLP:conf/chi/AdibMKKM15}
F.~Adib, H.~Mao, Z.~Kabelac, D.~Katabi, and R.~C. Miller, ``Smart homes that
  monitor breathing and heart rate,'' \emph{\emph{in} Proc. {ACM} {CHI}'15},
  pp. 837--846, 2015.

\bibitem{DBLP:journals/tmc/SiggSSJB14}
S.~Sigg, M.~Scholz, S.~Shi, Y.~Ji, and M.~Beigl, ``Rf-sensing of activities
  from non-cooperative subjects in device-free recognition systems using
  ambient and local signals,'' \emph{{IEEE} Trans. Mob. Comput.}, vol.~13,
  no.~4, pp. 907--920, 2014.

\bibitem{DBLP:conf/mobicom/WangZZWN14}
G.~Wang, Y.~Zou, Z.~Zhou, K.~Wu, and L.~M. Ni, ``We can hear you with wi-fi!''
  \emph{\emph{in} Proc. ACM MOBICOM'14}, pp. 593--604, 2014.

\bibitem{DBLP:conf/infocom/HanWWN14}
C.~Han, K.~Wu, Y.~Wang, and L.~M. Ni, ``Wifall: Device-free fall detection by
  wireless networks,'' \emph{\emph{in} Proc. {IEEE} {INFOCOM}'14}, pp.
  271--279, 2014.

\bibitem{DBLP:conf/sensys/HuangNG14}
D.~Huang, R.~Nandakumar, and S.~Gollakota, ``Feasibility and limits of wi-fi
  imaging,'' \emph{\emph{in} Proc. ACM SenSys '14}, pp. 266--279, 2014.

\bibitem{DBLP:conf/percom/MoellerRDKHOP}
A.~Moeller, L.~Roalter, S.~Diewald, M.~Kranz, N.~Hammerla, P.~Olivier, and
  T.~Ploetz, ``Gymskill: A personal trainer for physical exercises,''
  \emph{\emph{in} Proc. {IEEE} {PERCOM}'12}, pp. 588--595, 2012.

\bibitem{DBLP:conf/visapp/WangCCF14}
J.~M. Wang, H.~Chou, S.~Chen, and C.~Fuh, ``Image compensation for improving
  extraction of driver's facial features,'' \emph{\emph{in} Proc. {VISAPP}'14},
  pp. 329--338, 2014.

\bibitem{DBLP:conf/huc/SiggSJ13}
S.~Sigg, S.~Shi, and Y.~Ji, ``Rf-based device-free recognition of
  simultaneously conducted activities,'' \emph{\emph{in} Proc. ACM UbiComp
  '13}, pp. 531--540, 2013.

\bibitem{DBLP:conf/momm/SiggSBJW13}
S.~Sigg, S.~Shi, F.~B{\"{u}}sching, Y.~Ji, and L.~C. Wolf, ``Leveraging
  rf-channel fluctuation for activity recognition: Active and passive systems,
  continuous and rssi-based signal features,'' \emph{\emph{in} Proc. MoMM '13},
  p.~43, 2013.

\bibitem{HalperinAW:mobicom08}
D.~Halperin, T.~E. Anderson, and D.~Wetherall, ``Taking the sting out of
  carrier sense: interference cancellation for wireless {LAN}s,''
  \emph{\emph{in} Proc. ACM MOBICOM'08}, pp. 339--350, 2008.

\bibitem{DBLP:conf/pimrc/El-KafrawyYE11}
K.~El{-}Kafrawy, M.~Youssef, and A.~El{-}Keyi, ``Impact of the human motion on
  the variance of the received signal strength of wireless links,''
  \emph{\emph{in} Proc. {IEEE} {PIMRC}'11}, pp. 1208--1212, 2011.

\bibitem{DBLP:Polikar2006}
R.~Polikar, ``Ensemble based systems in decision making,'' \emph{IEEE Circuits
  and Systems Magazine}, vol.~6, no.~3, pp. 21--45, 2006.

\bibitem{DBLP:journals/tist/ChangL11}
C.~Chang and C.~Lin, ``{LIBSVM:} {A} library for support vector machines,''
  \emph{{ACM} {TIST}}, vol.~2, no.~3, p.~27, 2011.

\end{thebibliography}

\begin{IEEEbiography}[{\includegraphics[width=1in,height=1.25in,clip,keepaspectratio]{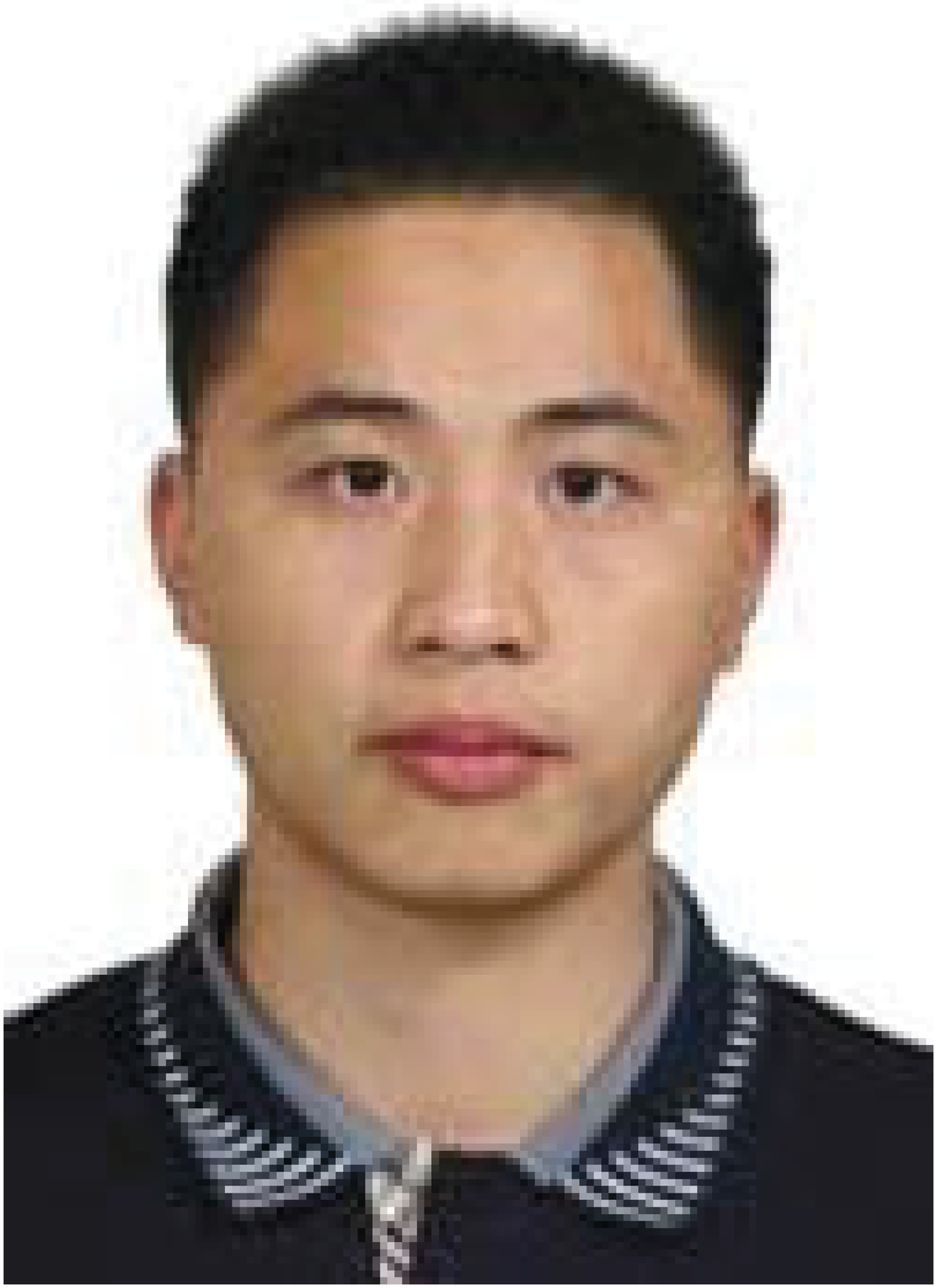}}]{Shaohe Lv} (S'6-M'11) is with the National Laboratory of Parallel and Distributed Processing, National University of Defense Technology, China, where he is an Assistant Professor since July, 2011. He obtained his Ph.D., M.D and B.S in 2011, 2005 and 2003 respectively, all in computer science. His current research focuses on wireless communication, machine learning and intelligent computing.
\end{IEEEbiography}

\begin{IEEEbiography}[{\includegraphics[width=1in,height=1.25in,clip,keepaspectratio]{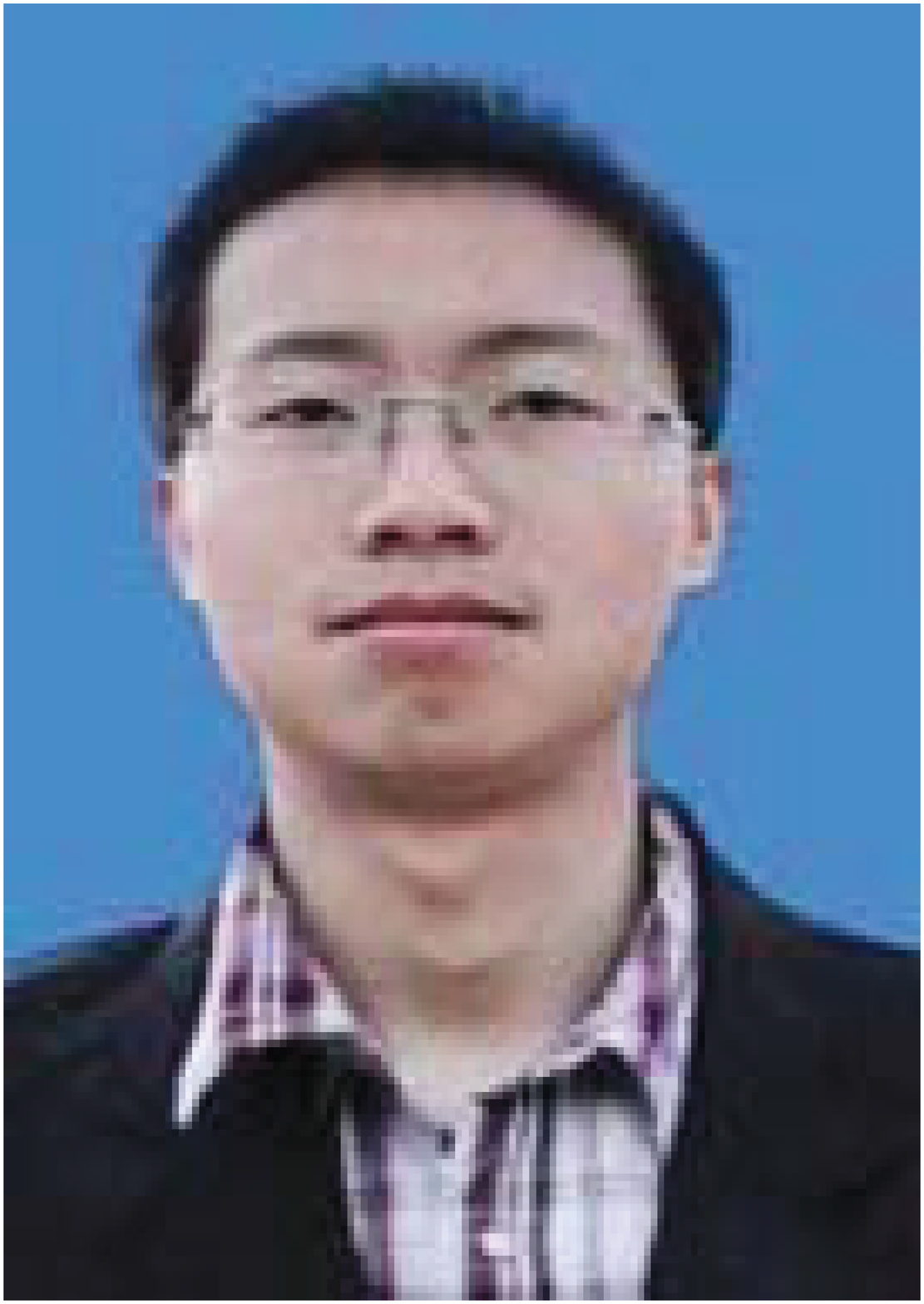}}]{Yong Lu} is with the National Laboratory for Parallel and Distributed Processing, National University of Defense Technology, China, where he is working towards a Ph.D. degree. His current research focuses on wireless communications and networks.
\end{IEEEbiography}

\begin{IEEEbiography}[{\includegraphics[width=1in,height=1.25in,clip,keepaspectratio]{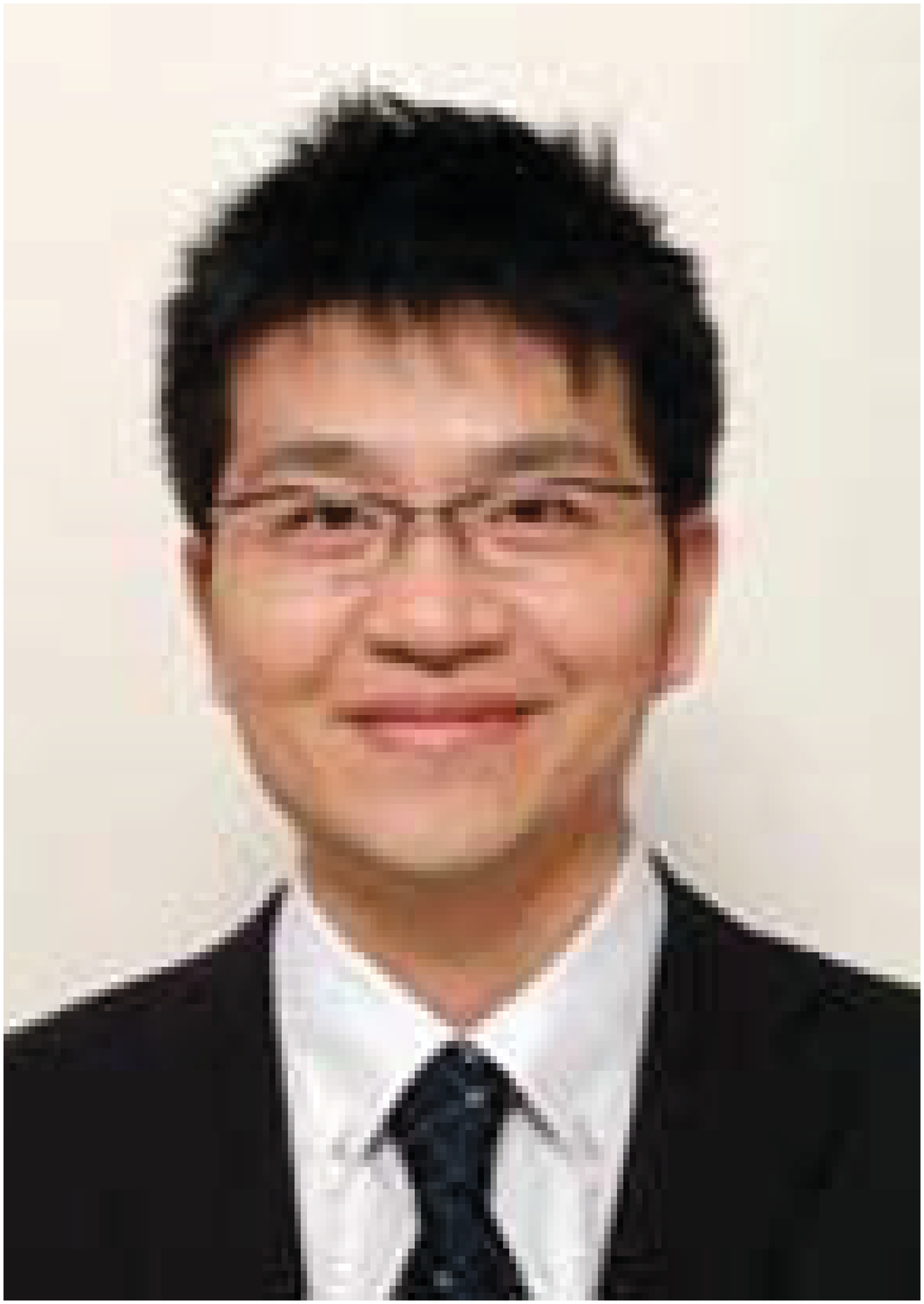}}]{Mianxiong Dong} is with the Department of Information and Electronic Engineering at the Muroran Institute of Technology, Japan where he is an Assistant Professor. He received his B.S., M.S. and Ph.D. in Computer Science and Engineering from The University of Aizu, Japan. His research interests include Wireless Networks, Cloud Computing, and Cyber-physical Systems. Dr. Dong is currently a research scientist with the A3 Foresight Program (2011-2016) funded by the Japan Society for the Promotion of Sciences (JSPS), NSFC of China, and NRF of Korea.
\end{IEEEbiography}

\begin{IEEEbiography}[{\includegraphics[width=1in,height=1.25in,clip,keepaspectratio]{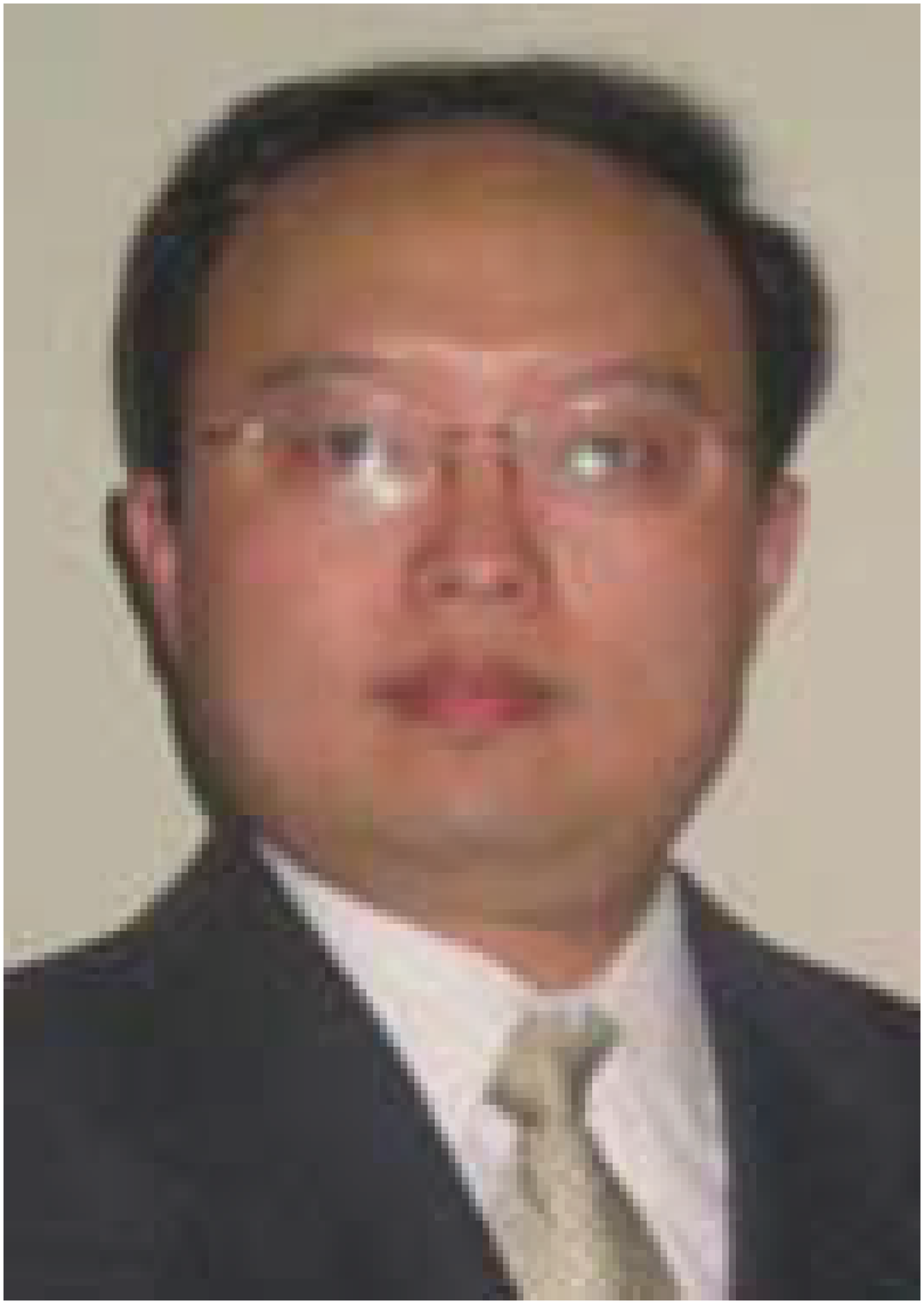}}]{Xiaodong Wang} is with the National Laboratory for Parallel and Distributed Processing, National University of Defense Technology, China, where he has been a Professor since 2011. He obtained his Ph.D., M.D and B.S in 2002, 1998 and 1996 respectively, all in computer science. His current research focuses on wireless communications and social networks.
\end{IEEEbiography}

\begin{IEEEbiography}[{\includegraphics[width=1in,height=1.25in,clip,keepaspectratio]{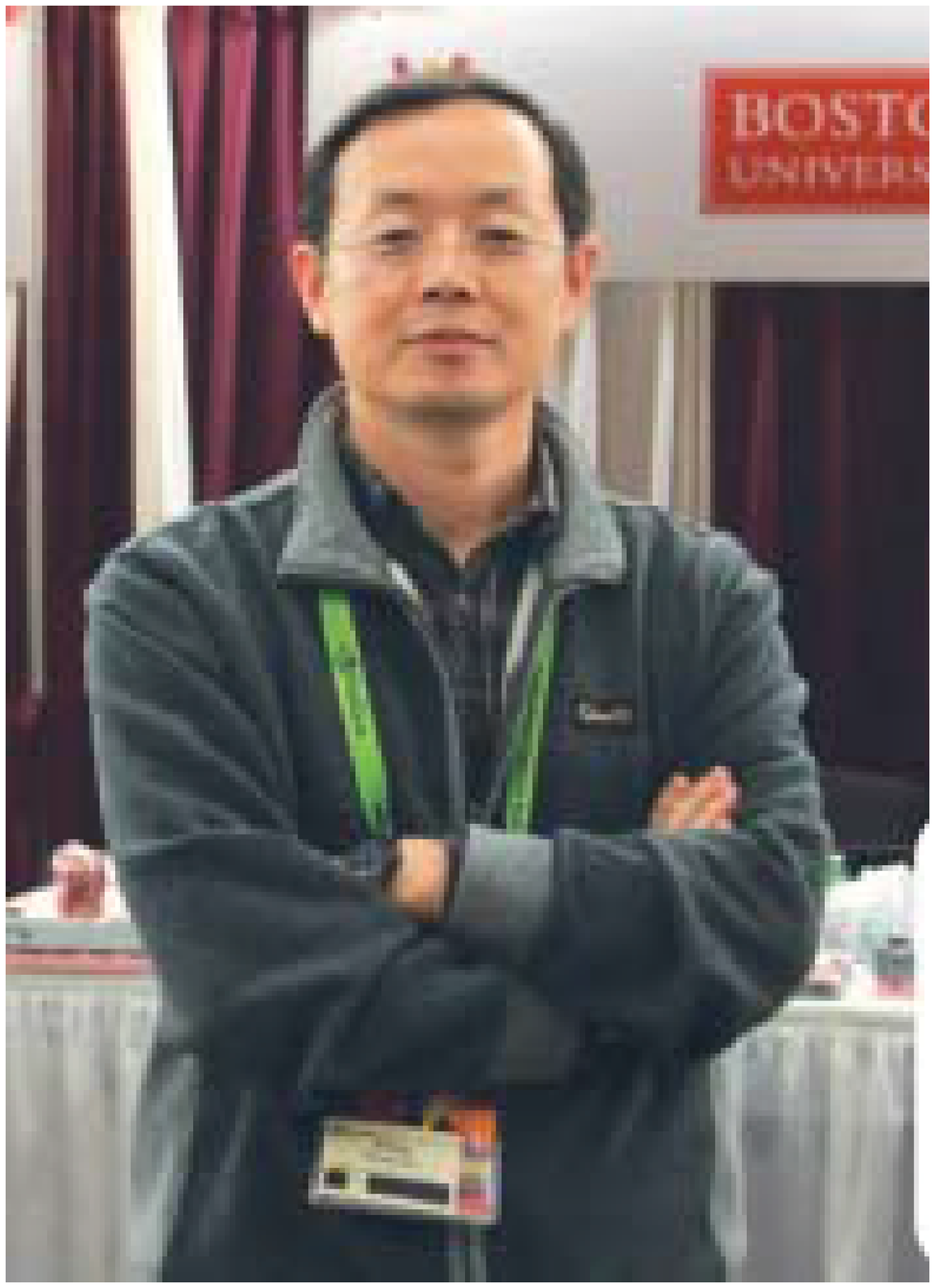}}]{Yong Dou} (M'08) is with the National Laboratory for Parallel and Distributed Processing, National University of Defense Technology, China, where he has been a Professor. His current research focuses on intelligent computing, machine learning and computer architecture.
\end{IEEEbiography}

\begin{IEEEbiography}[{\includegraphics[width=1in,height=1.25in,clip,keepaspectratio]{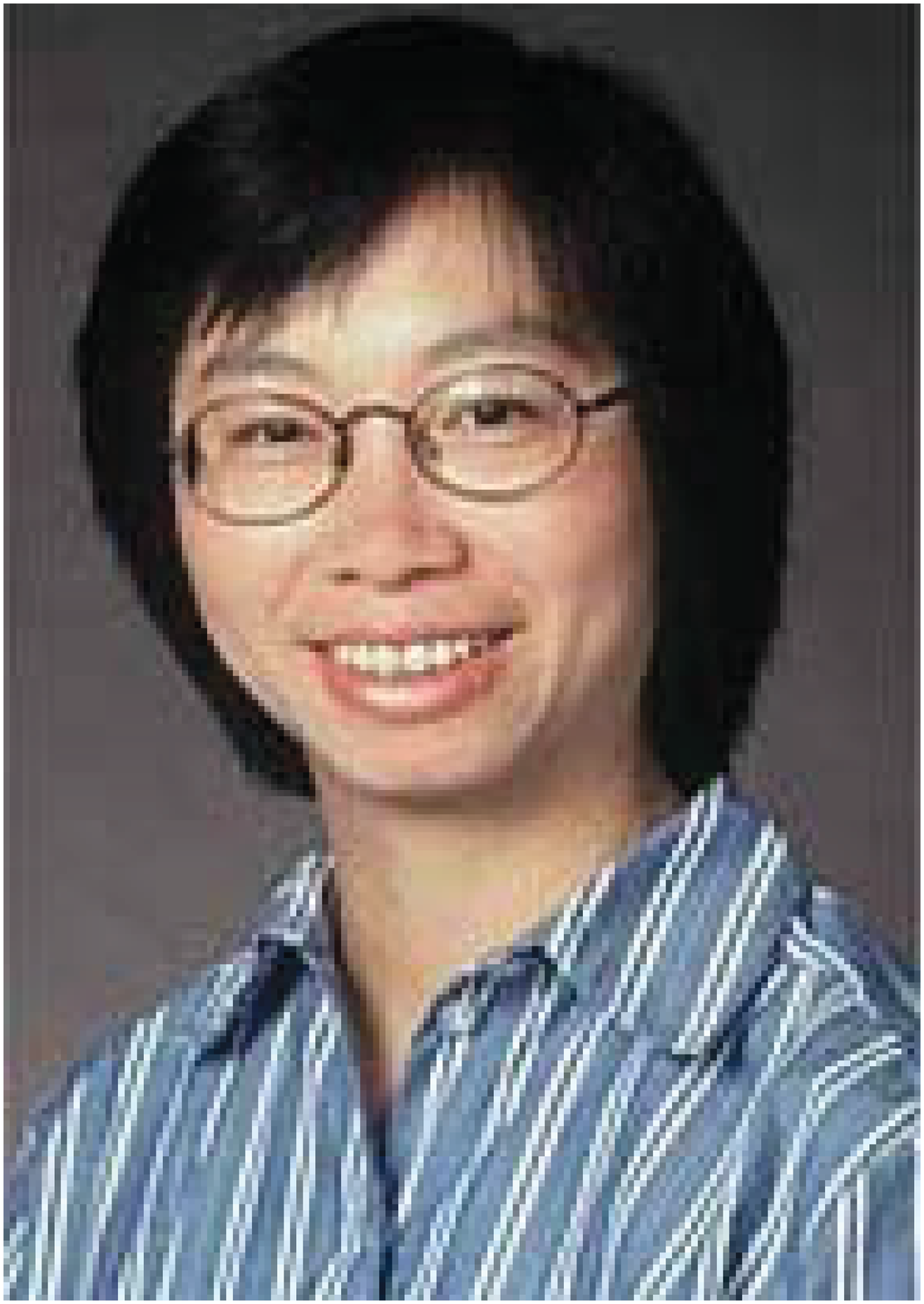}}]{Weihua Zhuang} (M'3-SM'01-F'08) is with the Department of Electrical and Computer Engineering, University of Waterloo, Canada, since 1993, where she is a Professor and a Tier I Canada Research Chair. Her current research focuses on wireless networks and smart grid. She is an elected member on the Board of Governors and VP Publications of the IEEE Vehicular Technology Society.
\end{IEEEbiography}

\end{document}